\documentclass[final,5p,times,twocolumn]{elsarticle}

\usepackage{subfigure}
\usepackage{graphicx}
\usepackage{ifpdf}
\ifpdf
\usepackage{epstopdf}
\fi
\usepackage{longtable}
\usepackage{amsmath}
\usepackage{color}
\usepackage[usenames,dvipsnames,svgnames,table]{xcolor}
\usepackage{lineno}
\usepackage{xspace}

\usepackage{bold-extra}
\usepackage{booktabs}
\usepackage{latexsym}
\usepackage{amssymb}
\usepackage{url}
\usepackage{epsfig}

\usepackage[ruled]{algorithm2e}
\usepackage{setspace}
\modulolinenumbers[5]

\newcommand{\Smilei}{{\sc Smilei}\xspace}

\newcommand{\vE}{\mathbf{E}}
\newcommand{\vB}{\mathbf{B}}
\newcommand{\vJ}{\mathbf{J}}
\newcommand{\vp}{\mathbf{p}}
\newcommand{\vx}{\mathbf{x}}
\newcommand{\vv}{\mathbf{v}}

\newcommand{\vF}{\mathbf{F}}

\begin{document}

\author[LLR]{A.~Beck \corref{mycorrespondingauthor} }
\cortext[correspondingauthor]{Corresponding author}
\ead{beck@llr.in2p3.fr}

\author[mdls]{J.~Derouillat}

\author[mdls]{M.~Lobet}
\ead{mathieu.lobet@cea.fr}

\author[intel]{A. Farjallah}

\author[LLR]{F.~Massimo}

\author[LLR]{I.~Zemzemi}

\author[LULI1,LULI2]{F.~Perez}

\author[LULI1,LULI2]{T.~Vinci}

\author[LULI1,LULI2]{M.~Grech}

\address[LLR]{Laboratoire Leprince-Ringuet, \'Ecole polytechnique, CNRS-IN2P3, F-91128 Palaiseau, France}
\address[LULI1]{LULI, CNRS, Ecole Polytechnique, CEA, F-91128 Palaiseau cedex, France}
\address[LULI2]{Sorbonne Univ., UPMC Univ. Paris 06, CNRS, LULI, place Jussieu, F-75252 Paris cedex 05, France}
\address[mdls]{Maison de la Simulation, CEA, CNRS, Universit\'e Paris-Sud, UVSQ, Universit\'e Paris-Saclay, F-91191 Gif-sur-Yvette, France}
\address[intel]{Intel Corporation, Meudon, France}

\title{Adaptive SIMD optimizations in particle-in-cell codes with fine-grain particle sorting}

\date{\today}

\begin{abstract}
Particle-In-Cell (PIC) codes are broadly applied to the kinetic simulation of plasmas, from laser-matter interaction to astrophysics.
Their heavy simulation cost can be mitigated by using the Single Instruction Multiple Data (SIMD) capibility, or vectorization, now available on most architectures.
This article details and discusses the vectorization strategy developed in the code \Smilei which takes advantage from an efficient, systematic, cell-based sorting of the particles.
The PIC operators on particles (projection, push, deposition) have been optimized to benefit from large SIMD vectors on both recent and older architectures.
The efficiency of these vectorized operations increases with the number of particles per cell (PPC), typically speeding up three-dimensional simulations by a factor 2 with 256 PPC.
Although this implementation shows acceleration from as few as 8 PPC, 
it can be slower than the scalar version in domains containing fewer PPC as usually observed in vectorization attempts.
This issue is overcome with an adaptive algorithm which switches locally between scalar (for few PPC) and vectorized operators (otherwise).
The newly implemented methods are benchmarked on three different, large-scale simulations considering configurations frequently studied with PIC codes.
\end{abstract}


\maketitle

\section{Introduction}

Particle-in-Cell (PIC) codes are among the most popular tools for the kinetic simulation of plasmas~\cite{birdsall}.
They consist in following the continuous trajectories of charged particles moving through a spatial domain under the action of external and self-induced electromagnetic (EM) fields.
These fields are represented on a discrete grid that also holds the plasma charge and current densities entirely defined by the particles' phase-space distribution.
Because the speed of light is finite, the nature of the physics described by EM PIC codes, the interactions between particles and fields, is spatially local.
This property makes them a good candidate for massive parallelization and several codes have indeed demonstrated
virtually unlimited weak scaling~\cite{smilei,osiris,picongpu} provided load balance is maintained~\cite{loadbalance}.
In contrast, the PIC algorithm is not well adapted for high performances at the single node level for several reasons.
First, in most cases, PIC simulations are becoming increasingly memory bound as memory performance is not ramping up as fast as the computation capabilities.
Second, particles are free to move anywhere in the domain and therefore trigger inefficient random accesses to memory each time they interact with the grid.
This randomness also prevents the use of Single Instruction Multiple Data (SIMD) instructions which are very efficient at speeding up memory-bound operations
but are restricted to very regular memory access patterns.
Finally, the wide variety of possible numerical configurations is difficult to optimize with a single technique.
PIC code optimization must therefore rely on randomness mitigation for optimized memory usage independently of the simulation parameters.

A first approach to mitigate randomness in PIC codes is to use the standard domain decomposition on domains so small that they can fit in the cache of the system.
This method is commonly used in recent implementations and the small domains are often referred to as \textit{patches}\cite{smilei} or \textit{tiles}\cite{vincenti2017,picador}
(from now on the term \textit{patch} is used as the generic denomination).
It exposes a very high level of parallelism and mitigates memory access randomness since particles of each patch all access
the same grid region which is limited by the patch extension.

Another approach to further reduce randomness is particle sorting.
This consists in organizing the particles in memory according to their location.
This idea was introduced in PIC codes in 1977 as part of a binary collision model \cite{Takizuka1977}, but only considered for optimization twenty years later on CPU  \cite{decyk1996,bowers2001} and on GPU a decade later \cite{gpusort,mertmann2011,decyk2014}.
Its purpose was to make memory accesses less random while maximizing the cache efficiency.
Since all computing systems have had multi-level caches for decades, sorting is nowadays very common in PIC codes, but it paradoxically implies a significant computation overhead because of potentially heavy data movements.
Moreover, when coupled to patching, sorting only has a minimal impact on cache management efficiency.
For those reasons, most PIC codes do not keep particles sorted at all times or perform only a coarse-grain sort\cite{gpusort}.
Nevertheless, in addition to cache use improvement, particle sorting can also favor SIMD operations by structuring memory accesses into repeatable patterns.
This article focuses on the fact that, with the increasing importance of these operations in today's hardware, the benefits of sorting at all times can actually overcome its cost.

The benefits of frequent sort and its possible implementation is discussed in \cite{nakashima2015,nakashima2017}.
In these works, particles are stored in many different cell-dependent arrays and moved in memory when they change cell.
This approach improves SIMD efficiency on
many-core architectures such as the Intel Xeon Phi provided that the particles arrays have enough elements.
A similar approach has been extended in \cite{barsamian_2018} where the authors use additional strategies such as the division of a cell's particle set into chunks to improve cache coherence and reduce memory transfers.
They report acceleration when using a few hundreds particles per cell.


The present work proposes a vectorized PIC algorithm based on a new fine-grain particle sorting at all times.
The algorithm relies on a cycle sort and retains a single particle array per patch.
It is combined with an adaptive mode that selects at runtime and locally (at the patch level) between the scalar and vectorized algorithms
depending on the local conditions in order to support efficiently any number of particles per cell.
This approach was implemented in the code \Smilei\footnote{\Smilei is an open-source project.
Both the code and its documentation are available online at \url{http://www.maisondelasimulation.fr/smilei}} \cite{smilei}, and its impact on the code performance is discussed throughout this paper.

The paper is structured as follows.
Section~\ref{sec:pic} summarizes the PIC algorithm and its implementation in \Smilei.
The performance of the most important operators acting onto the particles (namely the interpolator, pusher and projector), in their scalar version, is analyzed in terms of their computational cost.
The following Sec.~\ref{sort} details the fine-grain cycle sort algorithm and its benefits.
Section~\ref{sec:vectorization_operators} then focuses on the vectorization of each operator.
For generality, emphasis is placed on the algorithm rather than on the implementation itself.
Section~\ref{sec:vecto_efficiency} analyzes and compares the performance measurements between the scalar and vectorized operators.
We demonstrate that the vectorized algorithms are more efficient only for a large enough number of particles per cell.
This motivated the development of an adaptive method to select locally and dynamically (at runtime) the most efficient operators between scalar or vectorized
depending on local number of particles per cell in the patch.
This adaptive method is presented in Sec.~\ref{sec:adaptive_operators}.
Section \ref{sec:simulation_benchmark} presents the performance gain that can typically be obtained by using the fully vectorized and adaptive modes in large-scale 3D simulations.
Three configurations, two related to laser-plasma interaction the third one to astrophysics, are presented.
In all three cases, the scalar, vectorized and adaptive modes are used and their performances are compared.
Finally, conclusions are given in Sec.~\ref{sec:conclusion}.


\section{The PIC method}
\label{sec:pic}

This first section briefly summarizes the basics of the PIC method for collisionless plasma simulation.
This presentation introduces in particular the main operators that in \Smilei act onto the particles
and which performance, in their scalar version, will be presented at the end of the section.
More detailed descriptions of the PIC method can be found in \cite{birdsall,dawson1983,hockney1988},
and \Smilei's implementation is more specifically explained in \cite{smilei}.


\subsection{The Maxwell-Vlasov model}

The kinetic description of a collisionless (fully or partially ionized) plasma
relies on the Vlasov-Maxwell system of equations.
In this description, the different species of particles constituting the plasma are described by their respective distribution
functions $f_s(t,\vx,\vp)$, where $s$ denotes a given species consisting of particles with charge $q_s$ and mass $m_s$,
and $\vx$ and $\vp$ denote the position and momentum of a phase-space element.
The distribution $f_s$ satisfies Vlasov's equation\footnote{SI units are used throughout this work.}:
\begin{eqnarray}\label{eq_Vlasov}
\left(\partial_t  + \frac{\vp}{m_s \gamma} \cdot \nabla + \vF_L \cdot \nabla_{\vp} \right) f_s = 0\,,
\end{eqnarray}
where $\gamma = \sqrt{1+\vp^2/(m_s\,c)^2}$, $c$ is the speed of light in vacuum, and
\begin{eqnarray}\label{eq_LorentzForce}
\vF_L = q_s\,(\vE + \vv \times \vB)
\end{eqnarray}
 is the Lorentz force acting on a particle with velocity $\vv=\vp/(m_s\gamma)$.

This force follows from the existence, in the plasma, of collective electric [$\vE(t,\vx)$] and magnetic [$\vB(t,\vx)$] fields satisfying
Maxwell's equations:
\begin{subequations}\label{eq_Maxwell}
\begin{eqnarray}
\label{eq_BGauss} \nabla \cdot \vB &=& 0 \,,\\
\label{eq_Poisson} \nabla \cdot \vE &=& \rho/\epsilon_0 \,,\\
\label{eq_Ampere}\nabla \times \vB &=& \mu_0\, \vJ + \mu_0 \epsilon_0\,\partial_t \vE \,,\\
\label{eq_Faraday}\nabla \times \vE &=& -\partial_t \vB \,,
\end{eqnarray}
\end{subequations}
where $\epsilon_0$ and $\mu_0$ are the vacuum permittivity and permeability, respectively.

The Vlasov-Maxwell system of Eqs.~\eqref{eq_Vlasov} -- \eqref{eq_Maxwell} describes the self-consistent dynamics of the plasma
whose constituents are subject to the Lorentz force, and in turn modify the collective electric and magnetic fields through their charge
and current densities:
\begin{subequations}\label{eq_rhoJ}
\begin{eqnarray}
\rho(t,\vx) &=& \sum_s q_s\int\!d^3\!p f_s(t,\vx,\vp)\,,\\
\vJ(t,\vx) &=& \sum_s q_s\int\! d^3\!p\,\vv f_s(t,\vx,\vp)\,.
\end{eqnarray}
\end{subequations}

In the electromagnetic code \Smilei, velocities are normalized to $c$.
Charges and masses are normalized to $e$ and $m_e$, respectively, with $-e$ the electron charge and $m_e$ its mass.
Momenta and energies (and by extension temperatures) are then expressed in units of $m_e c$ and $m_e c^2$, respectively.
The normalization for time and space is not decided {\it a priori}.
Instead, all the simulation results may be scaled by an arbitrary factor, chosen here to be an angular frequency $\omega$. 
Temporal and spatial quantities are then expressed in units of $\omega^{-1}$ and $c/\omega$, respectively, 
while (number) densities are in units of $\epsilon_0 m_e \omega^2/e^2$.
More details are given in \cite{smilei}.

\subsection{Data structures: Macro-particles and fields}

The ``Particle-In-Cell'' method owes its name to the discretization of the distribution function $f_s$ as a sum of $N_s$ ``macro-particles''
(also referred to as ``super-particles'' or ``quasi-particles''):
\begin{eqnarray}\label{eq_fs_discretized}
f_s(t,\vx,\vp) = \sum_{p=1}^{N_s}\,w_p\,\,S\big(\vx-\vx_p(t)\big)\,\delta\big(\vp-\vp_p(t)\big)\,,
\end{eqnarray}
where $w_p$ is the $p^{th}$ macro-particle ``weight'', $\vx_p$ is its position, $\vp_p$ is its momentum.
$\delta(\vp)$ is the Dirac distribution
and $S(\vx)$ is the so-called shape-function of all macro-particles.
These macro-particles are advanced, knowing the electromagnetic fields at their position, by solving their relativistic equations of motion.
For convenience, in the rest of this article macro-particles will be referred to simply as ``particles''. \\
Particle weights, momentum components and position components are stored separately in contiguous arrays.
These arrays are elements of a structure of arrays called \verb+Particles+.
The EM fields experienced by the particles (obtained at the particles' positions after the interpolation step, see section \ref{interpolation})
as well as their Lorentz factors are stored in temporary contiguous arrays.

\begin{figure}
 \centering
 \includegraphics[width=0.48\textwidth]{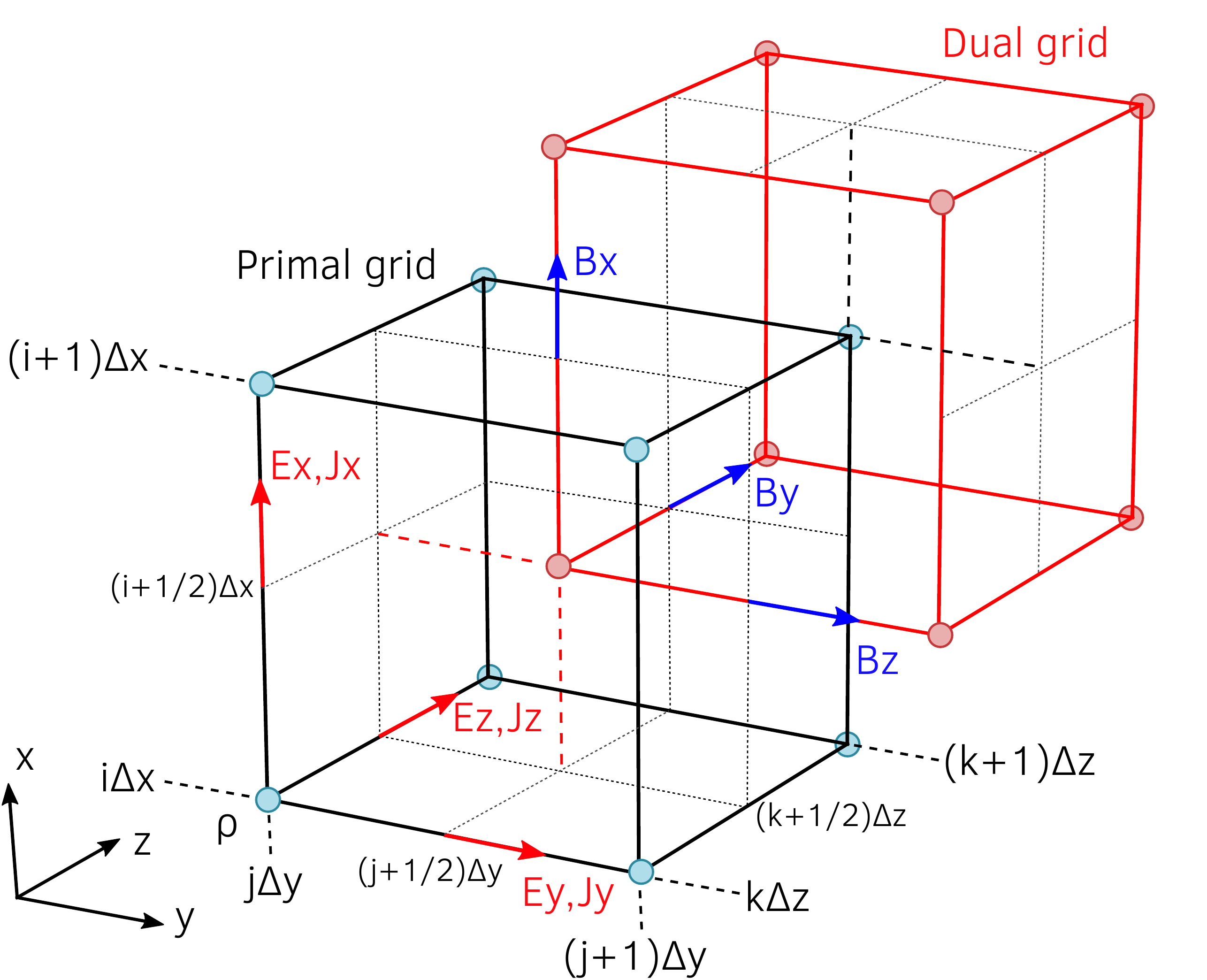}
 \caption{Representation of the staggered Yee grids. The location of all fields
 and current densities follows from the common convention to
 define charge densities at the cell vertices. The black cell is part of the
 \textit{primal} grid which vertices carry the charge density.
 The red cell is part of the \textit{dual} grid which vertices are located at the
 center of the primal cells. Primal and dual vertices are respectively represented
 by blue and red circles.}
 \label{fig_Yee}
\end{figure}

\Smilei uses the Finite Difference Time Domain (FDTD) method~\cite{taflove2005} to solve Maxwell's equations.
The EM field components, charge densities and current density components are thus stored onto Cartesian staggered grids as illustrated in Fig.~\ref{fig_Yee}.
This Yee grid~\cite{yee1966} is a very standard mesh layout used in most FDTD approaches,  as well as refined methods based on this technique~\cite{nuter2014}.
It involves two regularly-spaced grids: \textit{primal} and \textit{dual}. 
Primal vertices are points where the charge density $\rho$ is evaluated; they delimit the primal cells. 
Dual vertices are located at the center of the primal cells and form the dual grid.
Apart from the charge density, other quantities are not evaluated at either of these vertices, 
but at midpoints highlighted in figure \ref{fig_Yee}. 
As an example, the current component $J_x$ is dual in $x$, primal in $y$ and primal in $z$.

\subsection{The PIC time-loop iteration}

The explicit PIC time loop operations consist in solving successively
Maxwell's and Vlasov's equations.
Maxwell's equations are solved by performing an explicit FDTD solver.
Vlasov's equation, solved by advancing particles in phase space, requires three steps:
\begin{itemize}
    \item Field interpolation (or field gathering): the freshly updated electric
    and magnetic fields from the Maxwell solver, being known only at the grid vertices, are interpolated at each particle's position. 
    The interpolation method accounts for the fields of several neighboring cells according to a particle shape function.
    \item Particle push: the equation of
    motion is solved using the interpolated fields.
    This typically relies on finite difference leap-frog methods (e.g. schemes from Boris \cite{boris1970,birdsall}, Vay \cite{vay2008} or
    Higuera-Cary \cite{higuera2017}) which advance the momenta at
    the middle of the time step before computing the positions at the next time step.
    \item Projection (or current deposition): once the particles have been pushed,
    their contributions to the current need to be projected back to the grids.
    As field interpolation, this step uses the particle shape function.
    Note that, in \Smilei, current deposition relies on the charge-conserving method developed by Esirkepov~\cite{esirkepov},
    and this projection method will alway be considered throughout this work.
    The current projected onto the grids is then used in the Maxwell solver to
    compute the following time step.
\end{itemize}

\subsection{The PIC time-loop performance}
\label{sec:pic_scalar_performance}

In plasma simulations, advancing the particles is usually much more expensive than solving Maxwell's equations.
The computational cost thus scales with the number of particles which is, in most case, vastly larger than the number of vertices.
The computational cost also varies between operators.
In this Section, the performance of the scalar particle operators (namely the interpolator, pusher and projector) is analyzed.

To do so, we consider the simple case of a thermal plasma.
An homogeneous, Maxwellian, hydrogen plasma fills the entire simulation domain,
with an initial proton temperature of 10 keV and electron temperature of 100 keV,
and particles were initially randomly distributed in space.
The domain has periodic boundary conditions and the cell dimensions are
$\Delta x = \Delta y = \Delta z \simeq 0.22\ c / \omega$, where
$\omega$ denotes the electron plasma frequency in this particular case.
Simulations were run for 100 iterations with a time step $\Delta t = 0.95 \Delta_{\rm CFL} = 0.12 \omega^{-1}$
where $c \Delta_{\rm CFL} = \left( \Delta_x^{-2} + \Delta_y^{-2} + \Delta_z^{-2} \right)^{-1/2}$
corresponds to the timestep at the Courant-Friedrichs-Lewy (CFL) condition.
The shape function for interpolation and projection is of order 2, i.e. over 3 vertices in each direction,
and as stressed earlier Esirkepov's charge-conserving current projection scheme is used.

The simulations are performed on a single node of the Skylake
super-computer \textit{Irene Joliot-Curie} in France (see \ref{compilation}).
The domain is divided into $8\times8\times6$ patches.
The run has 2 MPI processes with 24 openMP threads each so that each core has 8 patches to handle.
Each patch contains $8\times8\times8$ cells, which is sufficiently small to have the field data in L2 cache.
The load is balanced during the entire simulation as the plasma remains uniform.
This study neglects the cost of communications between nodes, focusing instead on the particle operators (interpolator,
pusher, projector) and the Maxwell solver.
As a consequence, the type of particles and their velocities have little impact
on the results.

In the following, we present a parametric study of the scalar operators' performance as a function the number of particles per cell (from 1 to 256).
Throughout this work, the performance of various operators will be measured by the computation time per particle per iteration.
In order to facilitate the comparison between architectures, this computation time is considered at the node level.
More precisely, it is computed as:
\begin{eqnarray}\label{eq:comptime}
\tau_{\rm part} = \frac{T_{\rm wall-clock}}{N_{\rm part} \times N_t} \times N_{\rm Nodes}\,
\end{eqnarray}
where $T_{\rm wall-clock}$ is the wall-clock time spent in the considered operators, 
$N_{\rm part}$ is the total number of particles in the simulation, 
$N_t$ the number of timesteps over which the simulation is run
and $N_{\rm Nodes}$ is the number of nodes used for the simulation.
In Sec.~\ref{sec:simulation_benchmark}, we will also present a time-resolved version of this measure that is obtained by summing not over the total number of timesteps, 
but a reduced number of them and doing so several time during the simulation.

\begin{figure}
	\centering
	\includegraphics[width=0.49\textwidth]{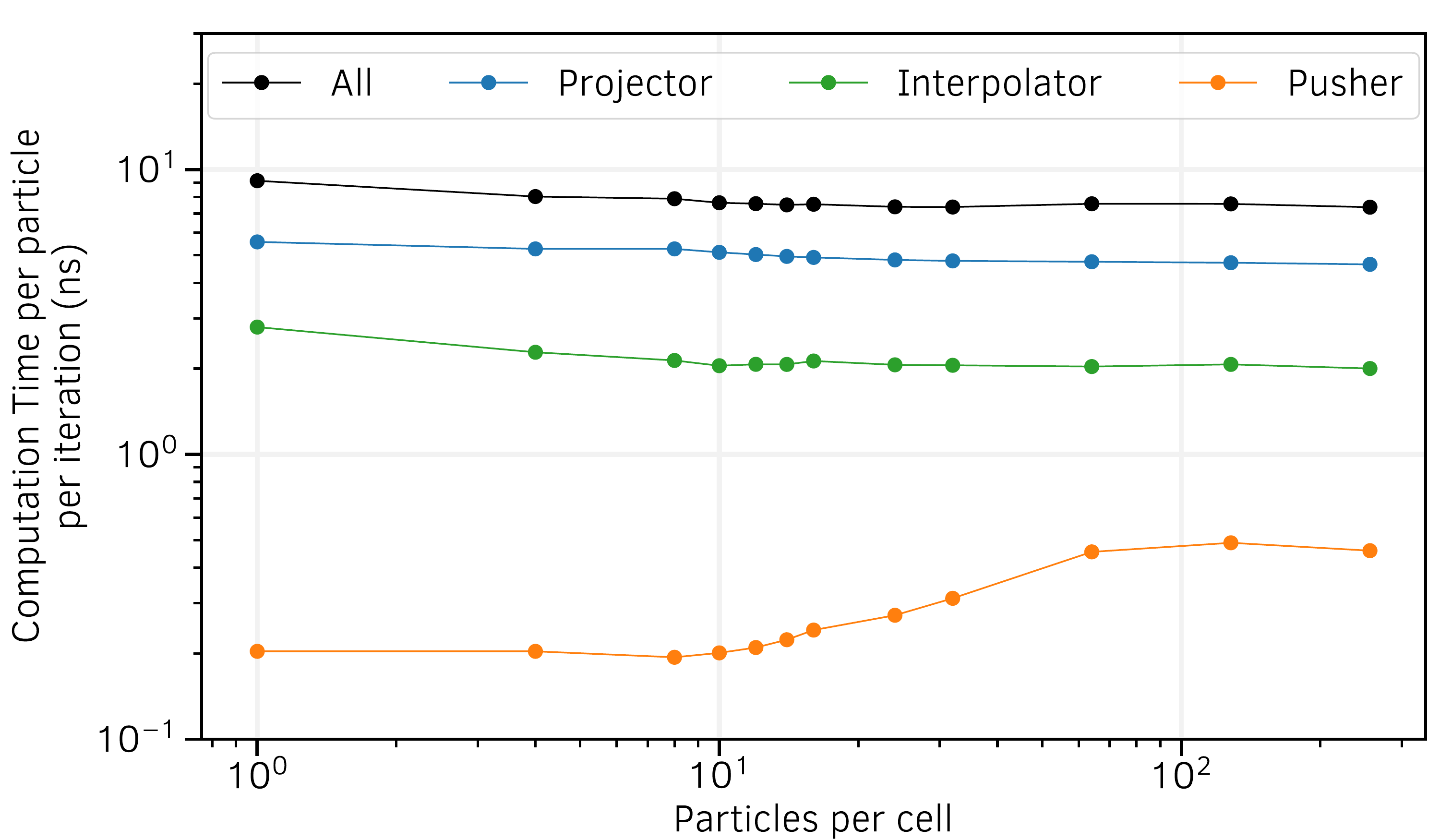}
	\caption{Computational cost [see Eq.~\ref{eq:comptime}] of each particle operator
		for the scalar version of the code, as a function of the number of particles per cell.
		Simulations run on a single Skylake node.}
	\label{fig_particle_scalar_operator_times_skl}
\end{figure}

The computation times obtained per particle and per iteration for each operator are shown in Fig. \ref{fig_particle_scalar_operator_times_skl}.
They appear to
depend weakly on the number of particles per cell, gaining $\sim$19\% at the higher end.
With little vectorization and neglecting cache issues, the scalar operators should
not depend on the number of particles per cell but on the total number of
particles to be computed per patch.
This is approximately verified.
The small gain is partly due to the scalar operators having vectorized sequences
(the code is always compiled with the vectorization flags)
even if the most intensive loops are not optimized.
Cache memory effects could also impact the particle computation time
but this analysis requires a deep instrumentation of the code.
The projection appears to be the most time-consuming operator ($\sim$65\% of the whole particle pushing time in average), followed by the interpolator ($\sim$30\% contribution). The pusher represents $\sim$5\% (or less) of the particle pushing time, although slightly increasing with the number of particles contrary to the other operators.
Note that the sum of all contributions is not exactly 100\% because the particle processing
includes additional small computation such as the exchange preprocessing.
The time spent in the Maxwell solver is independent of the number
of particles per cell and remains constant for all cases. In relative terms, it represents 12\% of the particle computation time for 1 particle per cell, and becomes rapidly negligible above, 
as particles consume more and more time.

The objective of the vectorization method described in this paper is to reduce the cost
of the three particle operators which are further detailed in sections \ref{interpolation}
 to \ref{projection}.


\section{Particle sorting}
\label{sort}

This section describes the algorithm used to sort particles in \Smilei which is a fine-grain and frequent sorting.

\subsection{Sorting definition and purpose}
A sorting technique in a PIC code is defined by: (i) the ``grain'' of the sorting, or resolution, often expressed as an elementary volume (i.e. sub-cells, single cell or multiple cells), (ii) the ordering of the set of grains, and
(iii) the frequency of the sorting (usually a fixed periodicity expressed in number of time steps).
The objective is that, after sorting, all particles within the same ``grain'' are stored contiguously in memory.

\begin{figure*}
	\centering
	\includegraphics[width=0.8\textwidth]{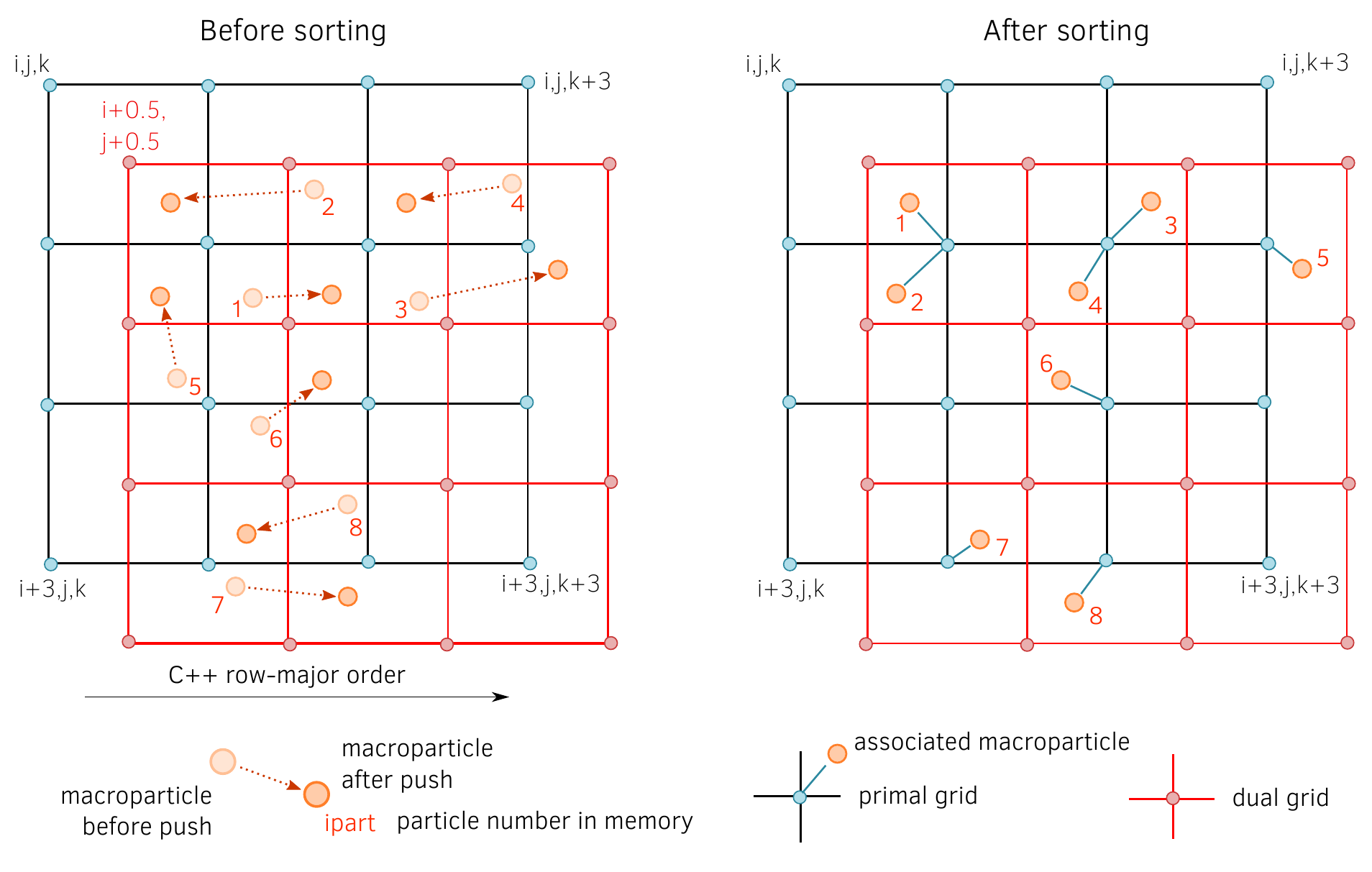}
	\caption{2D dual-cell-based particle sorting in \Smilei.
		The left panel represents the sorted particles before movements, and the unsorted ones afterwards.
		The right panel illustrates the ordered particles after sorting.
		Each panel outlines both primal and dual grid.
		Particles are sorted according to the nearest primal vertex (i.e. located in the same dual cell), the number next to each particle being its position in memory.
		After sorting, particles sharing the same primal vertex are contiguous in memory.}
	\label{sorting_scheme}
\end{figure*}

The vectorization strategy in \Smilei requires particles sharing the same primal indices to be contiguous in memory.
This is slightly different from a standard cell-based sorting and is in fact equivalent to a dual-cell-based sorting, as illustrated in figure \ref{sorting_scheme} for a two-dimensional situation.

Many authors suggest single-cell sortings are also a good practice to maximize cache efficiency.
Since it is usually not executed at every time step, the ordering of the sorted cells matters.
Indeed, cache use is optimized by the ordering if, as particles move, they only travel to cells close in memory to the cell they originate from.
It has been shown that ordering them along elaborate structures such as Z curves provides the best performances\cite{data-structure}.
However, the present the situation is different: the objective is to guarantee that SIMD operations can be executed at every time step.
Therefore the sorting, in \Smilei, must be done at every time step as well.
This high frequency ensures an optimized cache use, independently of the cell ordering.
As a consequence, a cell ordering that benefits best from the vectorized operators is chosen (see section \ref{projection}):
the \verb!C++! natural row-major order which matches that of all field data.

\subsection{Counting sort}
Sorting at every time step is a potentially costly operation which, without proper care, could overweight the benefits of having a well sorted array of particles.
The most expensive operation of the whole sorting process is particle copying because a single particle copy in memory involves a significant amount of data movement.
Consequently, an efficient sorting algorithm should aim at minimizing the number of particle copies.
In that regard, the counting sort has been a standard choice because it involves exactly one copy per particle.
The whole point of this algorithm is to determine, before any data movement, where each particle is supposed to be moved.
Pseudo code of the counting sort is given in algorithm \ref{countingsort} where the expression $range(N)$ refers to an array of integer ranging from 0 to $N-1$.

\begin{algorithm}
\DontPrintSemicolon
\KwData{
\\$Particles$: array of unsorted particles.
\\$CellKeys$: array of the cell indices of the particles.
\\$Count$: array counting the occurrence of each cell key.
\\$First\_index$: index of the first unsorted particle of each cell.
\\$Npart$: number of particles.
\\$Ncell$: number of cells.
}
\KwResult{$PartSorted$: array of sorted particles}
\Begin{
\textcolor{blue}{\tcp{$Count$ is evaluated.}}
\For{$ipart \in range(Npart)$}{
$Count[CellKeys[ipart]] \mathrel{+}= 1$\;
}
$First\_index[0] \longleftarrow 0$\;
\textcolor{blue}{\tcp{Accumulate $Count$.}}
\For{$icell \leftarrow 1$ \KwTo $Ncell-1$}{
$First\_index[icell] \longleftarrow First\_index[icell-1]+Count[icell-1]$\;
}
\textcolor{blue}{\tcp{Copy particles into the sorted array}}
\For{$ipart \in range(Npart)$}{
$PartSorted[First\_index[CellKeys[ipart]]] \longleftarrow Particles[ipart]$ \;
$First\_index[CellKeys[ipart]] \mathrel{+}= 1$\;
}
\KwRet  $PartSorted$

}
\caption{Counting Sort. }
\label{countingsort}
\end{algorithm}

This algorithm is standard in PIC codes where the sorting is executed at low frequency.
Between two sortings, each particle has time to travel several cells away from its original cell.
The algorithm must therefore be efficient at treating a completely disordered plasma and the counting sort is perfectly adapted to this.
Its major drawback is that it is an ``out-of-place'' sorting and therefore requires another full array of particles doubling the memory occupation of the particles.

\subsection{Cycling sort}

In the case of a high frequency sorting, there is little particle movement between two sortings and the particles remains relatively well ordered at all times.
Although these conditions seem favorable to the sorting, the counting sort still pays the full cost of one copy per particle and double memory.
It is therefore more efficient to use an ``in-place'' sorting
and copy only particles that effectively change cells.
This can be achieved with the cycle sort given in \ref{ap_cycle_sort}.
The purpose of this algorithm is to find a succession of circular permutations, or cycles, leading to a full sorted array while copying only particles which have effectively moved to a different cell.
Unlike the counting sort, the total number of copies is variable and depends on the particles movement and the length of the cycles found.
For a given cycle, the number of copies per particle is given by $N_c=(L+1)/L$ where $L$ is the length of the cycle.
This accounts for the necessary copy of one particle in a temporary variable.
The total number of particle copies can be approximated by $N_c\times N_m$ where $N_m$ is the total number of particles moving to a different cell.
In the worst case scenario, all cycles have the minimum length 2, $N_c=1.5$ and the number of copies is $1.5\times N_m$.
As long as $N_m < 2N_{part}/3$, where $N_{part}$ is the total number of particles, the total number of copies is still lower than when using a counting sort.
In general, few particles change cells between sortings when the sort is done frequently hence the obvious advantage of the cycle sort over the counting sort.

\subsection{Optimized cycle sort}

The cycling sort minimizes the number copies at the cost of a theoretical complexity of $\mathcal{O}\left(N_{part}^2\right)$:
for each particle at index $cycleStart$, the algorithm has to compute its future index in the array by traveling through all particles located after $cycleStart$.
This part of the algorithm can be significantly accelerated in the case of many duplicates.
This is usually the case in PIC codes because there are many more particles than cells and for that reason, many particles share the same $CellKeys$.
A useful optimization consists in building the $Count$ array in the same manner as in the counting sort.
This array is then used to keep track of the index where, in each cell, the next particle can be inserted.
It reduces the complexity of the algorithm to $\mathcal{O}\left(N_{part}+N_{cell}\right)$ so effectively to $\mathcal{O}\left(N_{part}\right)$ since $N_{part}\gg N_{cell}$ in most simulations.
The optimized cycling sort is given in algorithm \ref{optimizedcyclesort}.

\begin{algorithm}
\DontPrintSemicolon
\KwData{
\\$Particles$: array of unsorted particles.
\\$CellKeys$: array of the cell indices of the particles.
\\$Count$: array counting the occurrence of each cell key.
\\$First\_index$: index of the first unsorted particle of each cell.
\\$Last\_index$: index of the last particle of each cell.
\\$Npart$: number of particles.
\\$Ncell$: number of cells.
}
\KwResult{$Particles$: array of sorted particles}
\Begin{
    \textcolor{blue}{\tcp{$Count$ is initialized.}}
    \For{$ipart \in range(Npart)$}{
        $Count[CellKeys[ipart]] \mathrel{+}= 1$\;
    }
    $First\_index[0] \longleftarrow 0$\;
    \textcolor{blue}{\tcp{Accumulate $Count$.}}
    \For{$icell \leftarrow 1$ \KwTo $Ncell-1$}{
        $First\_index[icell] \longleftarrow First\_index[icell-1]+Count[icell-1]$\;
        $Last\_index[icell-1]\longleftarrow First\_index[icell]$\;
    }
    $Last\_index[Ncell-1]\longleftarrow Last\_index[Ncell-2]+Count[Ncell-1]$\;

    \textcolor{blue}{\tcp{Loop on each cell}}
    \For{$icell \in range(Ncell)$}{
    \For{$cycleStart\leftarrow First\_index[icell]$ \KwTo $ Last\_index[icell]$}{
     \If{$CellKeys[cycleStart]\mathrel{=}= icell$}{
         \textcolor{blue}{\tcp{Particle already well placed}}
         $\mathbf{continue}$\;
     }
         $cell\_dest \longleftarrow CellKeys[cycleStart]$\;
         $ip\_dest\longleftarrow First\_index[cell\_dest]$\;
         $Cycle.resize(0)$\;
         $Cycle.push\_back(cycleStart)$\;
         \textcolor{blue}{\tcp{Build a cycle}}
         \While{$ip\_dest\ \mathrel{!}=\ cycleStart$}{
             \textcolor{blue}{\tcp{Do not swap twins}}
             \While{$CellKeys[ip\_dest]\mathrel{=}=\ cell\_dest$}{
                 $ip\_dest \mathrel{+}= 1$\;
             }
             $First\_index[cell\_dest]\longleftarrow ip\_dest + 1$\;
             $Cycle.push\_back(ip\_dest)$\;
             $cell\_dest \longleftarrow CellKeys[ip\_dest]$\;
             $ip\_dest\longleftarrow First\_index[cell\_dest]$\;
         }
         \textcolor{blue}{\tcp{Proceed to the swap}}
         $Ptemp\longleftarrow Particles[Cycle.back()]$\;
         \For{$i \leftarrow Cycle.size()-1$ \KwTo $1$}{
             $Particles[Cycle[i]]\longleftarrow Particles[Cycle[i-1]]$\;
         }
         $Particles[Cycle[0]]\longleftarrow Ptemp$\;
       }
 }

\KwRet  $Particles$

}
\caption{Optimized Cycle Sort. }
\label{optimizedcyclesort}
\end{algorithm}

\subsection{Sorting in a parallel environment}
\label{sec:sorting-parallel}

\begin{figure}
	\centering
	\includegraphics[width=0.49\textwidth]{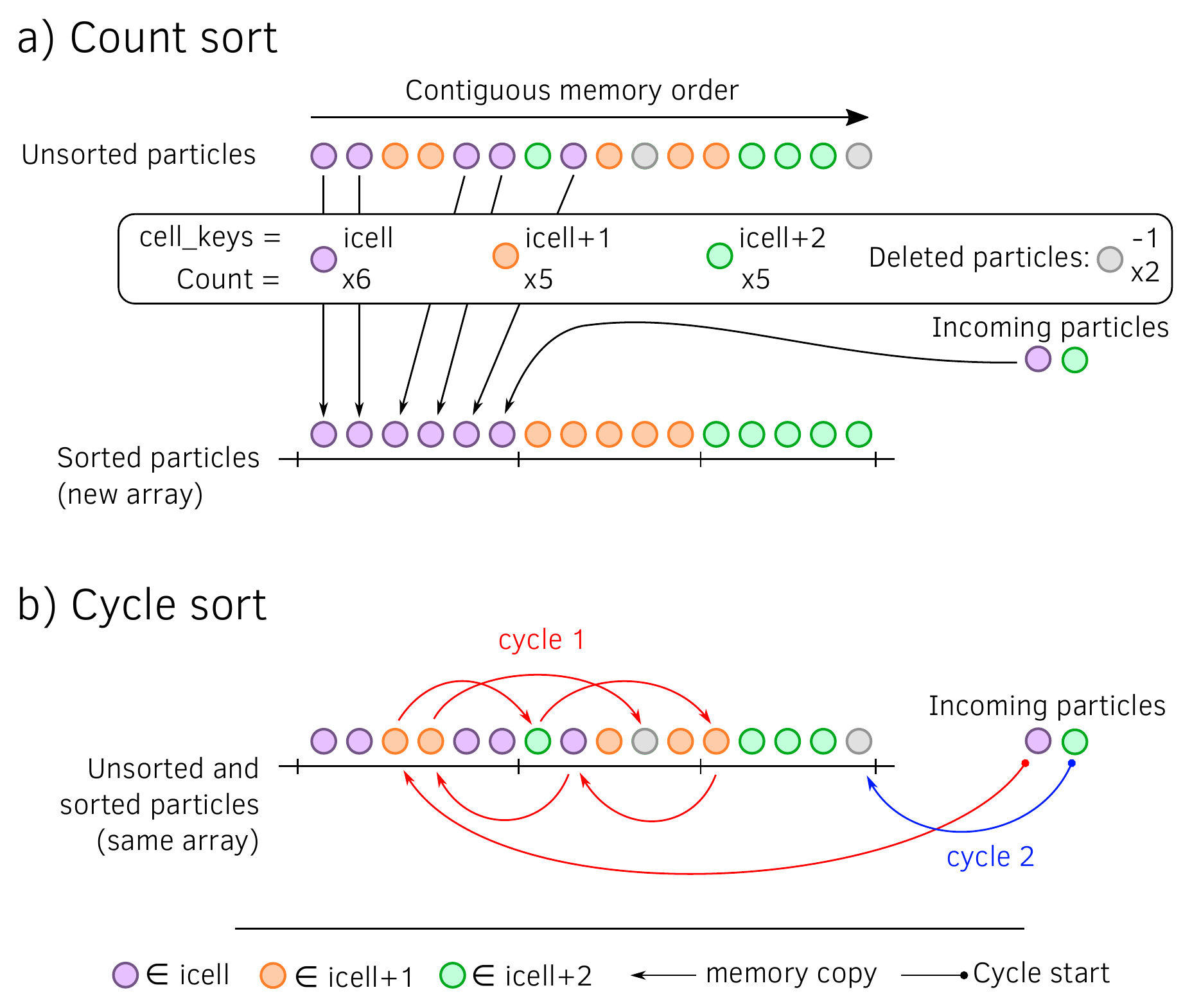}
	\caption{Comparison between counting sort and optimized cycle sort starting from an identical unsorted array of particles.
	Particles are colored as a function of their cell keys.
	Lost particles are given a cell key of -1.
	Particles coming from other patches are represented in a separate buffer.
	In panel a) the counting sort directly copies particles in a new array. Only copies required for the first cell are represented for readability.
	In panel b) the cycle sort performs particles permutations. Fewer copies are needed and they are all represented.}
	\label{sorting_sketch}
\end{figure}

PIC codes are usually executed in a parallel environment.
This poses two issues for the cycle sort algorithm.
First, particles are constantly exchanged with neighboring domains.
The size of each particle array changes and gaps appear in the middle of the array preventing a standard cycle sort.
A simple way of dealing with this issue is the following.
All particles entering in a given patch are stored in a buffer; they have their own $CellKeys$, and contribute to the $Count$ array.
All $CellKeys$ of exiting particles are set to $-1$.
The cycle sort algorithm is then executed through the particle array.
First cycles start with the entering particles and end when they hit a $CellKeys$ of $-1$.
The particles of these cycles are simply copied to their destinations, eventually overwriting the exiting particles.
This process is repeated for all entering particles.
At this point, all gaps are filled and all entering particles are placed.
If unsorted particles remain, the optimized cycle sort, from the previous section, is applied.
Figure \ref{sorting_sketch} sketches the whole process and illustrates the differences with the counting sort.

The second issue is related to load balancing.
Most advanced PIC codes deploy elaborate techniques in order to balance the computational load between the different compute units.
Since most of the computational load is proportional to the number of particles,
the effort mainly consists in balancing the number of particles per compute unit.
The counting sort cost is proportional to the number of particles as well and is not problematic.
However, the cost of a cycle sort strongly depends on the local disorder of the particle array, thus likely to cause significant load imbalance.
The disorder is difficult to estimate and taking it into account in a load balancing procedure proves complicated.
Instead, a good task scheduler can smooth this imbalance while being easier to achieve.
In \Smilei, this is the role of the OpenMP dynamic scheduler and the patch-based domain decomposition \cite{smilei}.


\section{Vectorization of the PIC operators}
\label{sec:vectorization_operators}

In most PIC simulations, an important fraction of the computation time is spent in the three main operators which are interpolation, pusher and projection.
This section describes the workflow of these 3 functions, how they are vectorized and why they benefit from the dual-cell fine-grain sorting.
Note that the vectorization effort in \Smilei focuses only on the algorithm and data structures and not on the implementation itself.
This means that no specific intrinsics were introduced in the code.
The only additions to the C++ code are \texttt{\#pragma omp simd} directives on critical loops and \texttt{aligned(64)} attributes to critical arrays.
Vectorization in \Smilei therefore relies only on auto-vectorization.

\subsection{Interpolation}\label{interpolation}

In the interpolation operation, also referred to as ``field gathering'', the EM field defined on the grid must be evaluated at each particle's position.
This operation on a single particle can be broken down into the three following sub-steps.
\begin{enumerate}
\item Extract the field data from the global field arrays in the neighborhood of the particle (the \textit{stencil}).
\item Compute the interpolation coefficients affecting each field data point, depending on the particle position relative to that of its cell.
\item Multiply fields by coefficients and sum all terms.
\end{enumerate}

\begin{figure}
 \centering
 \includegraphics[width=0.49\textwidth]{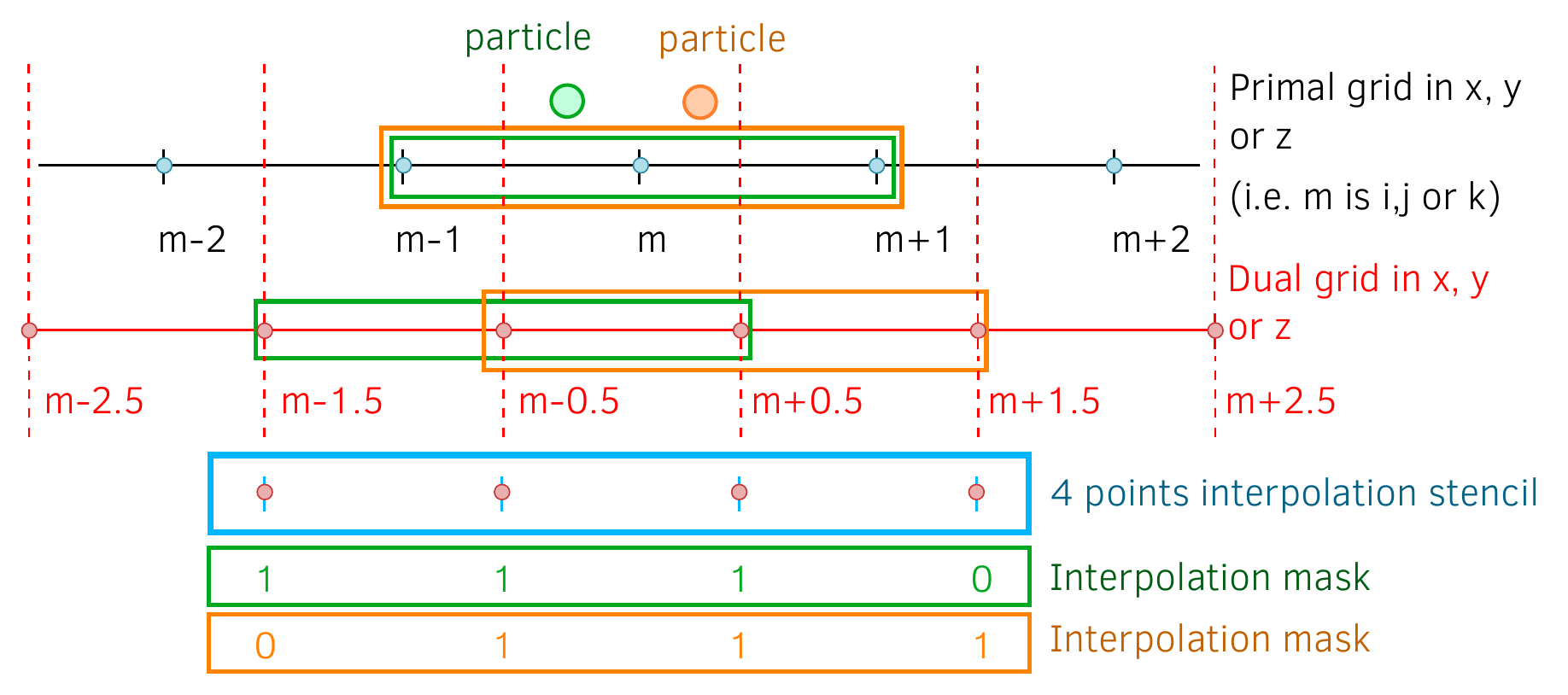}
 \caption{Illustration, in one dimension, of the primal and dual grid vertices accessed during the interpolation process with a second order shape function.
 Vertical dotted lines mark the boundary of the dual cells.
 Green and orange particles are in the same dual cell; they share the same primal index and therefore access the same 3 primal vertices.
 However, their dual indices differ by one which extends the number of dual vertices accessed to 4.}
 \label{fig_interpolation}
\end{figure}

The extracted portion of the field data depends on the position of the particle.
If particles are not well sorted, step 1 is a random access in a potentially large array.
In addition to significant cache misses, this also prevents SIMD operations.
It has even been reported that, for some architectures, the complicated pattern of interpolation behaved better when specifically instructing the compiler not to use SIMD operations \cite{picador}.
In \Smilei, as explained in section \ref{sort}, particles are sorted in dual cells so that groups of particles sharing the same primal indices are contiguous in memory.
These groups are treated successively and each of them is vectorized efficiently as follows.

The first benefit of sorting is that, for step 1, it completely removes particle dependency since all particles of the group require the same data. Access to the global memory is thus minimized and it improves cache use.
Primal components of the fields are common to all particles of the group since they share the same primal indices.
Dual components extend to only one additional vertex as illustrated on figure \ref{fig_interpolation}.
In these conditions, steps 2 and 3 can be easily vectorized.
They operate on the full stencil with the exception of one point depending on their initial position.
This is effectively dealt with via the use of a mask (see figure \ref{fig_interpolation}).

Sorting also guarantees that the local data involved in steps 2 and 3 are contiguous and can therefore be easily vectorized.
The local positions (relative to that of the cell) are stored for reuse later whereas the interpolation coefficients are loaded into a temporary buffer.
Particles groups are treated by sub-groups of 32 in order to limit the total size of these temporary buffers and fit them into the cache while retaining a reasonable vector length.
The optimal size of these sub-groups depends on the architecture and may change in the future.
Finally, the interpolated EM fields are returned and stored for each particle of the currently treated patch for later use in the pusher.


\subsection{Pusher}\label{pusher}

The pusher is the operation that benefits the most from vectorization with minimal
adjustments, provided that the data structure for particle properties is appropriate.
Its algorithm remains almost unchanged thanks to the optimized cycle sort.
It is performed on all particles of the patch regardless of the cell they occupy.

\begin{algorithm}
\DontPrintSemicolon
\KwData{
\\$particles$: array of sorted particles
}
\KwResult{$particles$: array of pushed and unsorted particles}
\Begin{

\textcolor{blue}{\tcp{Vectorized loop on particles}}
 \For{$particle \in particles$}{
   $\triangleright$ Update the momentum of $particle$\\
   $\triangleright$ Update the position of $particle$
 }


\KwRet  $particles$

}
\caption{Particle pusher.}
\label{particle_pusher}
\end{algorithm}

The \verb+CellKeys+ array, containing the dual cell index of each
particle, is updated after the pusher.
When a particle crosses the patch boundary,
\verb+CellKeys+ is set to $-1$ as a tag for the boundary condition treatment (see section \ref{sec:sorting-parallel}).


\subsection{Projection}\label{projection}


In the projection operation, also referred to as ``current deposition'', the
current density carried
by each particle must be evaluated at the coordinates of the surrounding vertices
and added to the current density global arrays.
Nowadays, one standard approach is the charge-conserving Esirkepov projection
algorithm \cite{esirkepov}.
The direct algorithm is not vectorizable in its naive form since two
particles located in the same cell could project their charge or current
contributions to the same vertices leading to memory races.
Nonetheless, efficiently vectorized algorithms have been implemented to get
around this limitation \cite{vincenti2017}.
Esirkepov's method is even more challenging to vectorize because
the computation not only depends on the particle's positions but also on their displacements.
Sorting suppresses all randomness in
positions but not in displacements.
This section explains how \Smilei benefits from sorting during the
projection phase and how it deals with the displacements randomness.

In Esirkepov's projection method, the current densities along each dimension
of the grid are computed from the charge flux through the cell borders.
By definition, fluxes are computed from the particle present and former
positions, respectively $x^{t}$ and $x^{t-\Delta t}$.
The simpler direct projection algorithm only uses the present particle position
$x^{t}$, but does not conserve charge.


In both methods, the operation on a single particle can be broken down
into the two following sub-steps:
\begin{itemize}
    \item Step 1 - Compute the projection contributions of the particle depending on its
    relative position in its local cell and its displacement.
    \item Step 2 - Add these contributions to the global array of current density
    according to the particle global position.
\end{itemize}

The Esirkepov projection, operating with a shape function of either 2nd or 4th order, has been vectorized by exploiting
 the properties of the sorted particles.
Algorithm \ref{particle_projection} presents concisely the method at 2nd order.
This algorithm is repeated for each current component $J_x$, $J_y$, $J_z$.

\begin{algorithm}
	\DontPrintSemicolon
	\KwData{
		\\$clusters$: List of clusters of 4 cells
		\\$vectors$: List of vectors of 8 particles contained in a given cell cluster
		\\$J_{local}$: local buffer to gather the current contribution from the cluster particles
	}
	\KwResult{$J$: current grids}
	\Begin{
		
		$\triangleright$ For each current component $J_x$, $J_y$ and $J_z$:
		
		\textcolor{blue}{\tcp{Loop 1 - on 4-cell clusters}}
		\For{$cluster \in clusters$}{
			
			\textcolor{blue}{\tcp{Loop 2 - on the cluster cells}}
			\For{$cell \in cluster$}{
				
				\textcolor{blue}{\tcp{Loop 3 - on particle vectors}}
				\For{$vector\in vectors$}{
					
					\textcolor{blue}{\tcp{Loop 4 - Vectorized loop on the vector particles}}
					\For{$particle \in vector$}{
						$\triangleright$ Compute each particle coefficients and distances to the vertices \\
						$\triangleright$ Compute each particle charge weight
					}
					
					\textcolor{blue}{\tcp{Loop 5 - Vectorized loop on the vector particles}}
					\For{$particle \in vector$}{
						$\triangleright$ Compute the current contributions and store in $J_{local}$
					}
					
				}
				
			}
			
			\textcolor{blue}{\tcp{Loops 6 - on vertex indexes}}
			\For{$i \in range(5)$}{
				\For{$j \in range(5)$}{
					\textcolor{blue}{\tcp{Vectorized loop in the $z$ contiguous direction}}
					\For{$k \in range(8)$}{
						\textcolor{blue}{\tcp{Unrolled loop on the particle vector size}}
						\For{$ipart \in range(8)$}{
							$\triangleright$ Reduction of $J_{local}$ in the main current array $J$
						}
					}
				}
			}
		}
	}
	\caption{Particle projection for order 2.}
	\label{particle_projection}
\end{algorithm}


In order to take advantage of the sorting, the first loops (1 and 2 in
Algorithm \ref{particle_projection}) iterate on the cells.
The particularity of \Smilei's projection
is the gathering of the cells in clusters of 4 cells in the
 $z$ direction (of index $k$).
Note that the number of cells per cluster actually depends on the order of the projection and the size of 4 cells is only valid for order 2.
The advantages of this decomposition is clarified in the following description.

\begin{figure}
	\centering
	\includegraphics[width=0.49\textwidth]{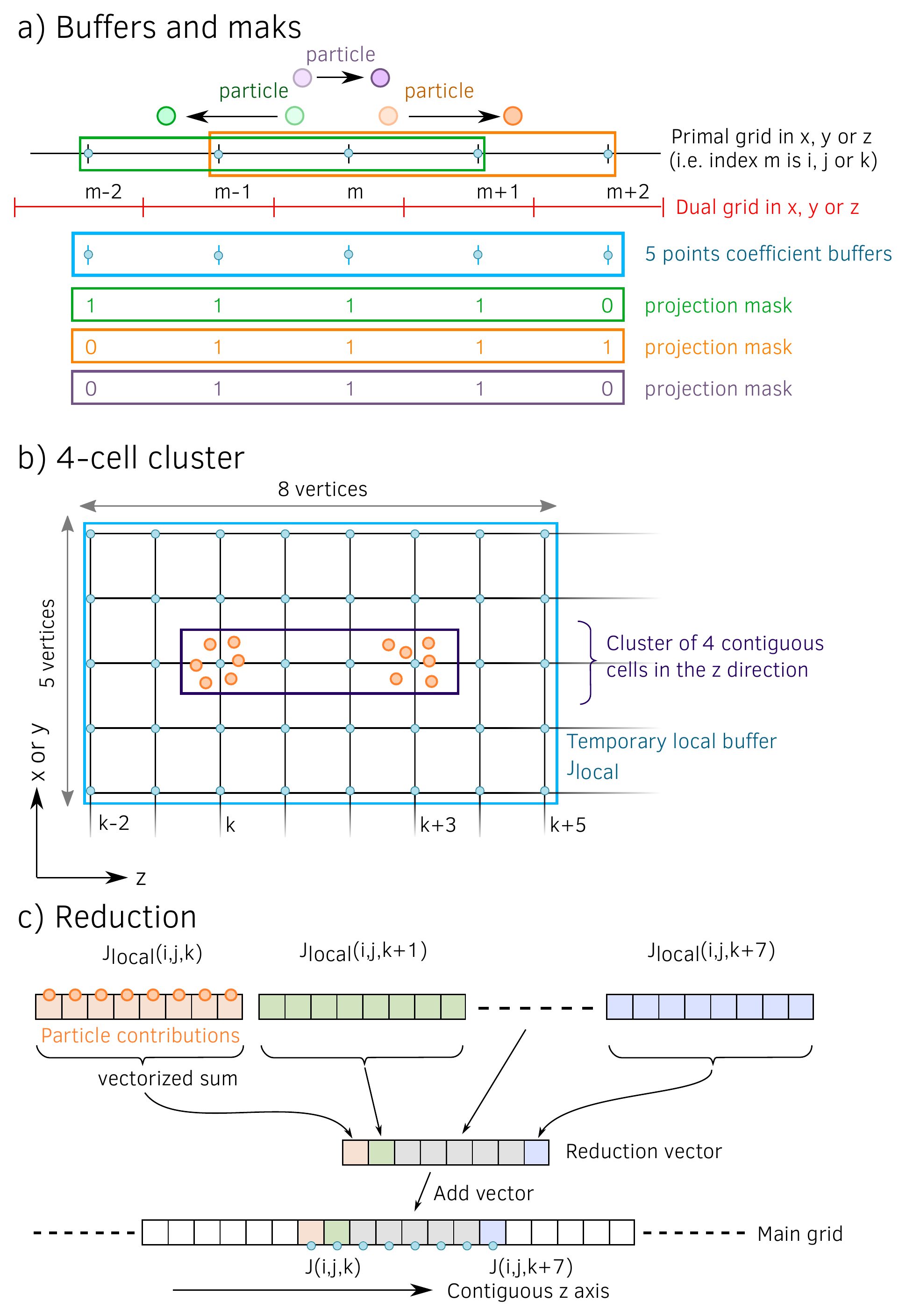}
	\caption{a) Primal vertices accessed during the projection process with
		a second order shape factor.
		b) Schematic of the multi-cell approach that uses
		a larger temporary current projection buffer in the z direction and
		helps reducing the number of projections in the patch grids.
		c) Drawing of the $J_{local}$ local buffer reduction process into the main array $J$.}
	\label{fig_projection}
\end{figure}

The first part of the algorithm corresponds to step 1.
Particles in the same dual cell (sharing the same primal indices)
are clustered into vectors of 8 particles to minimize the local buffer size while retaining enough data for an efficient vectorization.
The loop on these vectors is denoted \textit{loop 3} in Algorithm \ref{particle_projection}.
As illustrated in Fig. \ref{fig_projection}a, computing the coefficients
requires up to 4 vertices for each particle in each direction but potentially 5 vertices
if all particles of the same cell are considered.
This is due to the Esirkepov scheme which applies a shift depending on the particle displacement.
Since each particle only uses 4 vertices among 5, it has one
useless value at one vertex.
For the vectorization, the shape factor coefficients and
the flux intermediate coefficients are computed and stored in separate and
adequate buffers for each direction.
These buffers are carefully allocated (aligned and contiguous in the particle
direction) so that the computation of these coefficients is vectorized in the
particle loop 4 in Algorithm \ref{particle_projection}.
Each buffer has by default a size of 8 (vector particles) $\times$ 5 (vertices).


The computation of step 2
can be divided into 2 sub-steps.
During sub-step 2.1 (loop 5 in Algorithm \ref{particle_projection}), the current contributions,
calculated using previously computed coefficients, are stored on a small local grid, called $J_{local}$
in Algorithm \ref{particle_projection},
in order to avoid concurrent memory access and enable vectorization.
$J_{local}$ is treated like eight small, separate current grids so that
the current of the 8 particles of each vector can be stored independently
without concurrency.
In 3D, the grid size required to satisfy the projection of the particles in the 4-cell
cluster is $5 \times 5 \times 8$ cells as schematically shown in \ref{fig_projection}b.
Therefore, the local buffer $J_{local}$ is composed of $5 \times 5 \times 8 \times 8$ elements (12.5 kB).
The fast (contiguous) axis is the particle index.


Sub-step 2.2 (loop 6 in Algorithm \ref{particle_projection}) reduces the local grid $J_{local}$ into the main one $J$.
Vectorization is applied on the direction $z$, contiguous for $J$.
The 4-cell cluster enables to have 8 elements in this direction.
For each vertex, the 8 particles' contributions to $J_{local}$ are
summed in a temporary buffer as described in Fig. \ref{fig_projection}c.
This buffer is then added to the main grid $J$.
The 4-cell cluster further contributes to optimize this step by pooling 4 reductions.



The particle vectors size and the number of cells in a cluster may be adjusted to optimize the vectorization efficiency.
A large vector size requires more memory that will not necessary be used entirely if
there are not enough particles per cell.
The buffer memory size needs to be low enough to fit in L2 cache.
Larger cell clusters may help minimizing the number of reductions
but requires more memory.

When the number of particles per cell is not a multiple of the vector size (8
for \textsc{AVX512}, 4 for \textsc{AVX2}),
the remaining particles are treated in a smaller vector.
When the number of cells in $z$ is not a multiple of the cluster size (4), the
remaining cells are treated sequentially (i.e. one reduction per cell).


\section{Vectorization performances}
\label{sec:vecto_efficiency}



The vectorized operators implemented in \Smilei are designed to be efficient
when a systematic sorting algorithm is used, as described above.
Their performance is first evaluated using the 3D homogeneous Maxwellian
benchmark from section \ref{sec:pic_scalar_performance}, as a function of the
number of particles per cell (PPC) ranging from 1 to 256.
This study is focused on the particle operators (interpolator,
pusher, projector, sorting) and discards the computational costs of the Maxwell solver and of the communications between processes.
The patch size is kept constant at $8\times8\times8$ cells.


The test runs have been performed on 4 clusters equipped with different Intel
architectures typically used for \Smilei: Haswell, Broadwell,
Knights Landing (KNL) and Skylake.
The clusters' properties and the code compilation parameters are described in \ref{compilation}.
Each run has been performed on a single node.
Since the number of cores varies from an architecture to another, the runs were conducted
so that the load per core (i.e. OpenMP thread) is constant.
In other words, the number of patches per core is constant.
The total number of patches for each architecture is determined so that
each core has 8 patches to handle.
There is 1 MPI process per NUMA domain (NUMA stands for non-uniform memory access)
which means a single process per socket on Haswell,
Broadwell and Skylake nodes that all have 2 sockets per node.
A KNL node, configured in quadrant cache mode, has only 1 socket, and among the 68 available cores, 64 are used for the simulations and 4 for the system.
The total number of patches is of $8\times4\times3$ on Haswell (24 cores),
$8\times8\times4$ on Broadwell (32 cores), $8\times8\times8$ on KNL (64 cores), $8\times8\times6$ on Skylake (48 cores).

\begin{figure}
 \centering
 \includegraphics[width=0.49\textwidth]{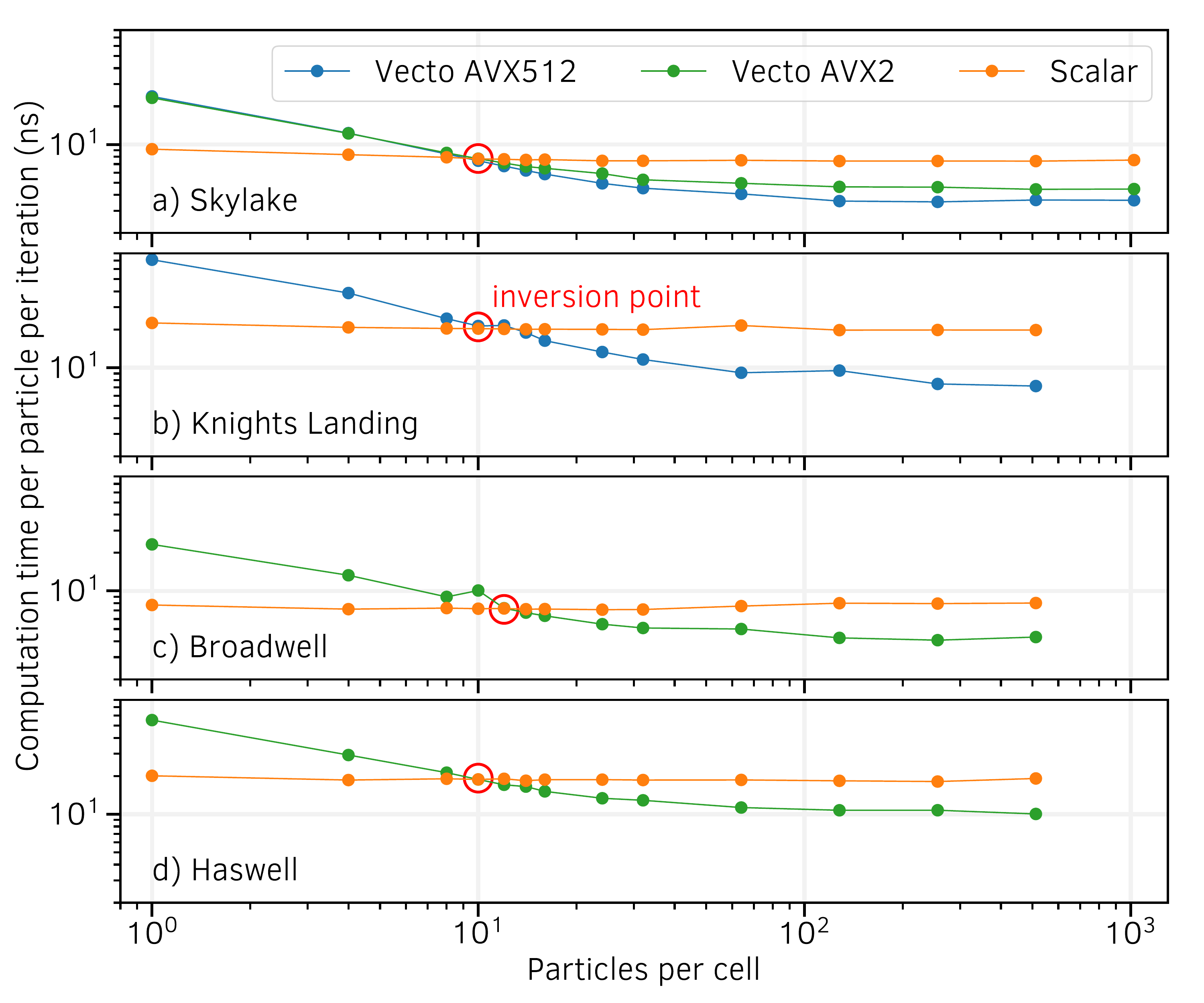}
 \caption{Particle computational cost
 as a function of the number of PPC.
 Vectorized operators are compared to their scalar versions on various cluster architectures.
 Note that the Skylake compilations accepts both AVX512 and AVX2 instruction sets.}
\label{fig_particle_times}
\end{figure}

The first series of tests considers an interpolation shape function of order 2
and compares the computation times
to advance a particle (interpolation, pusher, projection) per iteration.
The results for both scalar and vectorized versions are shown in Fig. \ref{fig_particle_times}.
Contrary to the scalar mode, the vectorized operators efficiency depends strongly on the number of PPC.
It shows improved efficiency, compared to the scalar mode, above a certain number of PPC denoted ``inversion point`` in Fig. \ref{fig_particle_times}.

The lower performances of the vectorized operators at low PPC can be easily understood.
First, their complexity is higher than their scalar counter-parts.
As explained in sec. \ref{sec:vectorization_operators}, the interpolation and projection masks increase the arithmetic intensity of the operations on a single particle.
Moreover, there are two additional loops, one over the cells and one sub-loop over groups of particles.
They are ineffective if cells are practically empty of particles.
And finally, SIMD instructions operate at a lower clock frequency than scalar ones \cite{Intel2018}.
For low numbers of PPC, these overheads are not compensated by the more efficient SIMD operations because the vector registers are not entirely filled and do not provide enough gain.

The location of the inversion point depends on the architecture:
10 PPC for Haswell and Broadwell, 12 for KNL, and 10 for Skylake,
considering the most advanced instruction set for each processor type.
Since Skylake can handle both the \textsc{AVX512} and the \textsc{AVX2}
instruction sets, the results from the two compilations are presented
in Fig. \ref{fig_particle_times}a for comparison.
The compilation in \textsc{AVX2} does not affect the run performance below the inversion
point when the scalar mode dominates.
However, the \textsc{AVX512} mode appears up to 30\% more efficient  than \textsc{AVX2} above 10 PPC.

In vectorized mode, the computation time decreases
with the number of PPC and stabilizes after 100 PPC around a final value that depends on the architecture.
On Haswell, the efficiency gains a factor of 1.9 at 512 PPC compared to the scalar mode.
On Broadwell, the same value is reached at 256 PPC.
On KNL, a factor of 2.8 is obtained at 512 PPC
(the highest for all considered architectures), but this fills the entire high-bandwidth memory (16 Gb), preventing tests above.
On Skylake, a maximum gain of 2.1 is reached at 256 PPC with \textsc{AVX512}, while reaching 1.7 at 1024 PPC with \textsc{AVX2}.

\begin{figure}
 \centering
 \includegraphics[width=0.49\textwidth]{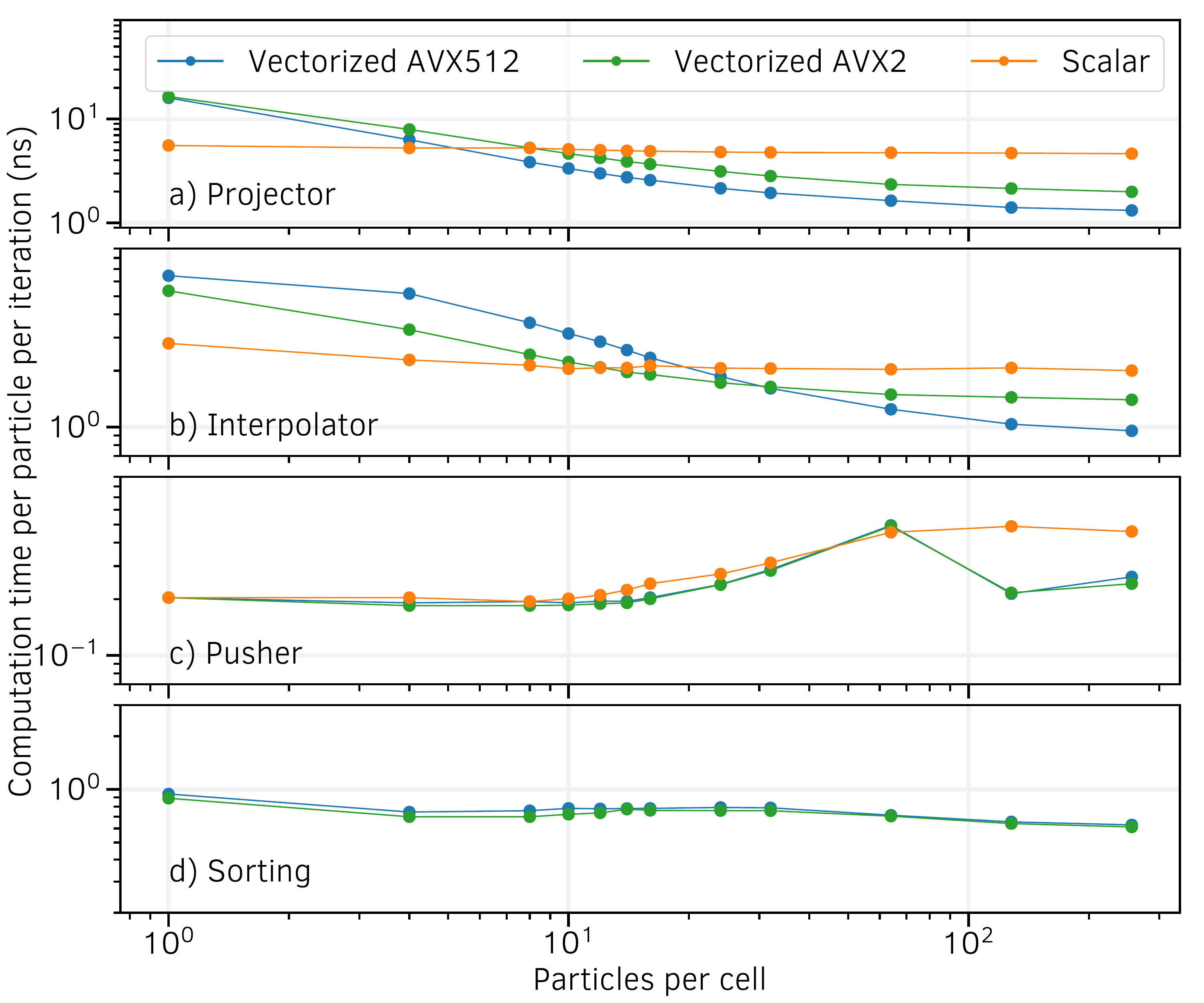}
 \caption{Computational cost of the four particle operators as a function of the number of PPC, for the Skylake test cases, in vectorized and scalar modes.
 Scalar cases are always compiled with the most
 advanced instruction set.}
\label{fig_particle_operators}
\end{figure}

Neglecting memory and cache effects, an ideal vectorization should give
an almost constant computation time per particle when the vector registers are filled.
In other words, the maximum gain from vectorization should be equal to the vector register size (8 in double precision on the most recent architectures, KNL and Skylake, when compiled with the AVX512 instruction set,
and 4 in double precision with the AVX2 instruction set for Haswell and Broadwell).
As demonstrated above, \Smilei's vectorized algorithms are not perfect due to
the nature of the operators (interpolation and projection) that induce the presence
of semi-vectorized or scalar sequences.
This is highlighted in Fig. \ref{fig_particle_operators} showing the computational cost of each operator from the same test case, simulated on Skylake only.
With \textsc{AVX512} vectorization, the projector remains the most time-consuming operator, even though it features the highest gain compared to scalar mode: $\times$3.5 at 256 PPC (and $\times$2.5 with \textsc{AVX2}).
Its cost decreases from 66\% of the total particle time at 1 PPC to 37\% at 256 PPC.
The interpolator is most efficient above 32 PPC with \textsc{AVX512} vectorization reaching a speed factor of 2 compared to the scalar mode, and 1.5 compared to \textsc{AVX2}.
The pusher remains negligible: it represents  1 to 12\% of the total particle time, depending on the number of PPC.
As it is automatically vectorized with the compilation flag \textsc{-03}, there is little speed gain. However, it benefits from the decomposition of particle data into blocks, thus showing higher efficiency above 128 PPC.
The cost of the sorting operation does not depend much on the number of PPC (although, in relative terms, it varies from 1 to 18\% of the total particle cost).
Indeed, the complexity of the cycle sort would only increase with a higher proportion of particles changing cells (higher temperatures or higher $\Delta t/\Delta x$), which does not vary with the number of PPC in a given thermal plasma benchmark.
This step does not benefit from vectorization (mainly data transfer), thus there is therefore no
difference between \textsc{AVX2} and \textsc{AVX512}. However, its cost is low compared to the speed gain from the interpolator and the projector optimizations, consequently ensuring an overall improvement.

The same parametric study is performed with different electron temperatures $T_e$ ranging
from 10 keV to 100 MeV and an ion temperature of $T_i = T_e / 10$.
The results confirm that the sorting cost increases with the temperature
whereas the interpolation, projection and pusher cost are unchanged.
At 100 MeV, the sorting takes 7 \% longer than the interpolator.
At this temperature, the thermal velocity (most probable velocity) is close to $c$ and
more than half of the particles change cell every time step.
These are extreme conditions for EM PIC codes but yet, the cost of the sorting is still
compensated by the vectorization speed-up.


The same parametric study has been conducted with a 4th-order interpolation shape function.
The global trends are similar to those at order 2:
in scalar mode, times do not depend significantly on the number of PPC, while they decrease in vectorized mode.
The inversion point is located at 10 PPC for Haswell and Broadwell, 4 for KNL and 6 for Skylake.
At 256 PPC, the vectorized particle operators (\textsc{AVX512}) are
respectively 1.4 faster on Haswell, 1.7 on Broadwell , 5 on KNL and 2.8 on Skylake , compared to the scalar version.
The most recent architectures benefit the most from vectorization, in particular with KNL which may prove even faster with more PPC.


\section{Adaptive Vectorization Mode}
\label{sec:adaptive_operators}



According to section \ref{sec:vecto_efficiency}, the scalar operators are significantly
 more efficient when the number of PPC is under the inversion point, which depends on the architecture.
However, in both laser-matter interaction or astrophysical cases, the number of PPC may be vastly different from one domain to another, and this number may evolve significantly during a simulation.
Consequently, the vectorized (or scalar) operators may not be adequate in all spatial regions, or for all times.
This issue can be addressed by using an adaptive vectorization mode which can locally switch
between the scalar and vectorized operators during the simulation, choosing the most efficient one in the region of interest.
Every given number of time steps, for each patch, and for each species, the most efficient operator is determined from the number of PPC.
This provides an automated, fine-grain adjustment in both space and time.
It also contributes
to the dynamic load balancing since patches with more PPC will be treated more efficiently.
This mode is now referred to as ``adaptive''.
Note that, two different adaptive modes exist in \Smilei:
\begin{itemize}
\item Adaptive mode 1: the sorting methods of the scalar and vectorized operators
are different, respectively the standard coarse-grain sort and the cycle sort
described in section \ref{sort}.
Switching modes thus requires sorting particles again.
\item Adaptive mode 2: the cycle sort method is used with both operators.
The scalar operators have been adapted to fit the new sorted structure.
\end{itemize}

A naive criterion to determine which operators should be applied locally
consists on using a threshold on the average number of particles per
cell.
Another simple method, implemented at first, consists on counting the number of
cells with particles below and above the inversion point.
Then, the ratio of the two is computed.
A threshold on this ratio determines the most suitable operators.
With a statistical study, an adequate threshold could be found although the
criterion was still proposing the wrong operators when the
particle distribution was broad.
This criterion appears nonetheless computationally cheap and satisfying in many cases.

A more complex empirical criterion has been developed.
It is computed from the parametric studies presented
in \ref{sec:vecto_efficiency}. Fig. \ref{fig_particle_times} summarizes their results
and indicates, for a given species in a given patch, the approximate time to compute the particle
operators using both the scalar and the vectorized operators.
The computation times have been normalized to that of the scalar operator for a single particle and 2nd-order shape functions.
The outcomes from different architectures appear sufficiently similar to consider an average between their results, as shown in the same figure.
\begin{figure}
 \centering
 \includegraphics[width=0.49\textwidth]{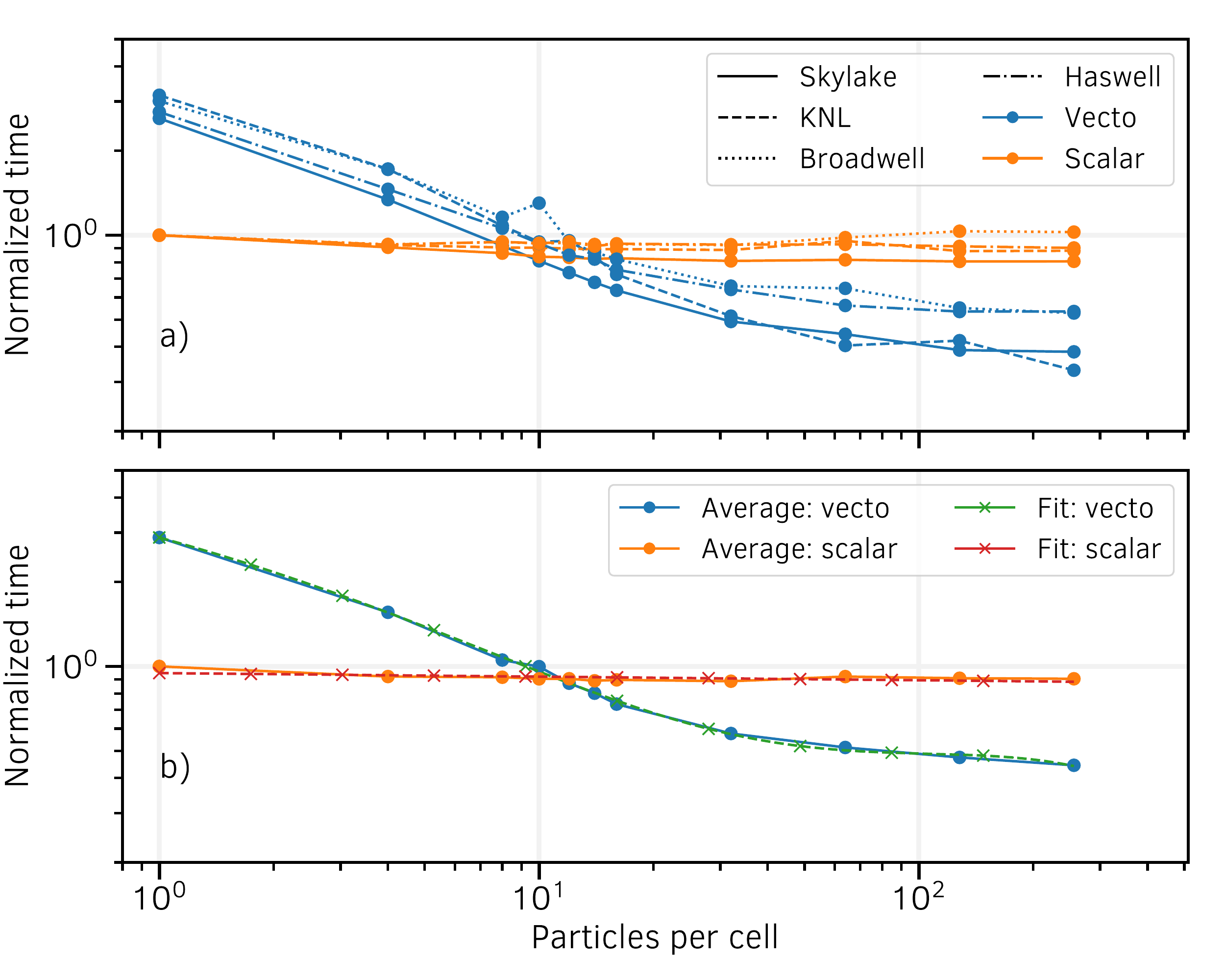}
 \caption{a) Normalized time per particle spent for all
 particle operators in the scalar and vectorized modes with various architectures, and 2nd-order interpolation shape functions.
 b) Averages of the curves in panel a), and polynomial regressions. }
\label{fig_vecto_efficiency_o2_all_fit}
\end{figure}
A linear regression of the average between all the scalar results writes
\begin{eqnarray}
S(N) = -1.17 \times 10^{-2} \log{\left( N \right)} + 9.47 \times 10^{-1}
\end{eqnarray}
where $S$ is the computation time per particle normalized to that with 1 PPC, and $N$ is the number of PPC.
For the average between vectorized results, a fourth-order polynomial regression writes
\begin{eqnarray}
 V(N) = -4.27 \times 10^{ -3 } \log{ \left( N \right)}^4 \\ \nonumber
+ 3.69 \times 10^{ -2 } \log{ \left( N \right)}^3 \\ \nonumber
+ 4.07 \times 10^{ -2 } \log{ \left( N \right)}^2 \\ \nonumber
 -1.07 \log{ \left( N \right) } \\ \nonumber
+ 2.88
\end{eqnarray}
These functions are implemented in the code to determine approximately the normalized single-particle cost.
Assuming every particle takes the same amount of time, the total time to advance a species in a given patch can then be simply evaluated
with a sum on all cells within the patch as
\begin{equation}
\sum_{c \ \in\ patch\ cells} N(c) \times F\!\left(N(c)\right)
\end{equation}
where $F$ is either $S$ or $V$.
Comparing these two total times, for $S$ and $V$, determines which of
the scalar or vectorized operators should be locally selected. This operation is repeated every given number of time steps to adapt to the evolving plasma distribution.
Note that similar approximations may be computed for specific processors instead of using a general rule.
In \Smilei, other typical processors have been included, requiring an additional compilation flag.

\begin{figure}
 \centering
 \includegraphics[width=0.49\textwidth]{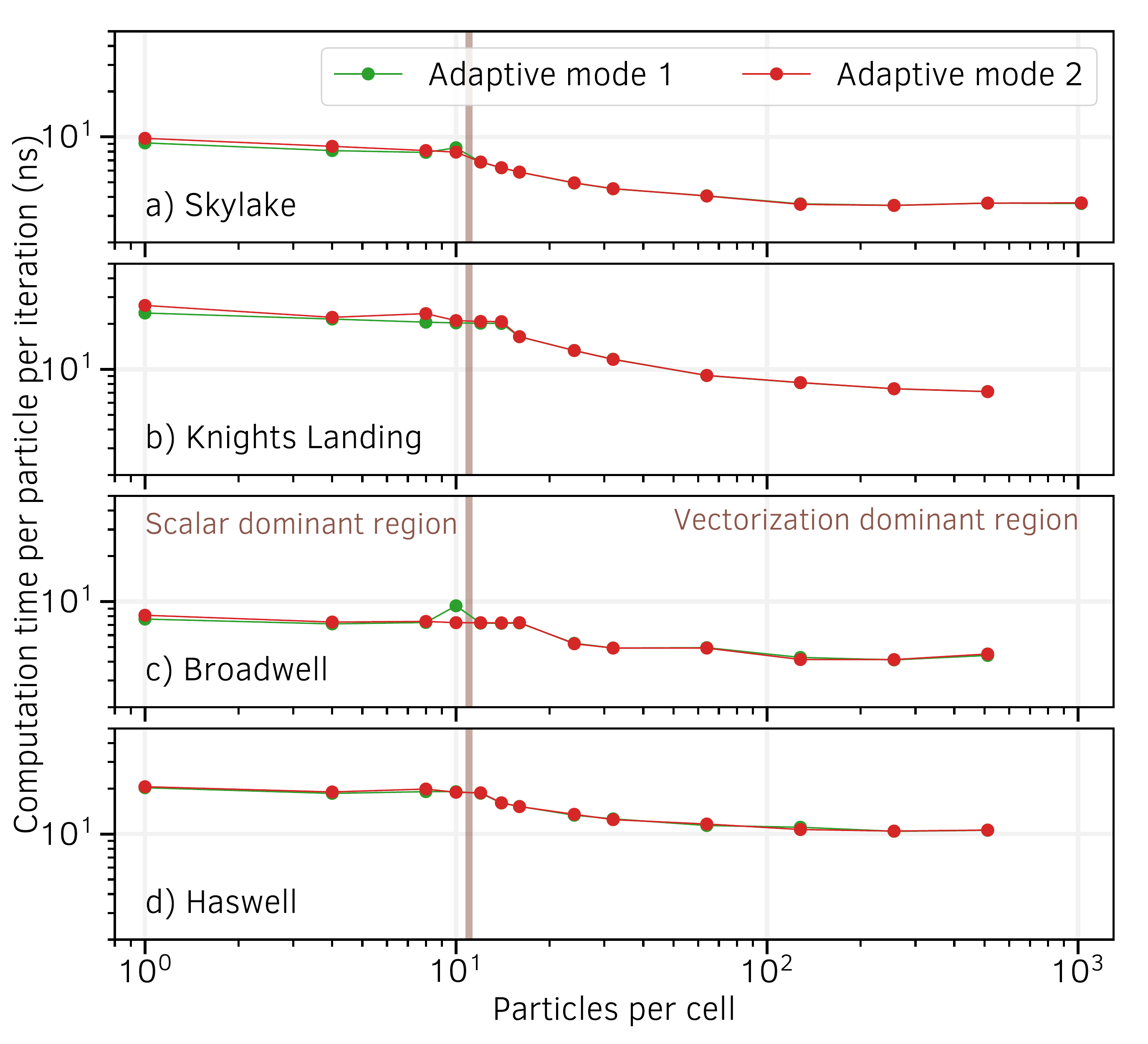}
 \caption{Particle computation times as a function of the number of PPC in the adaptive modes 1 and 2, for various architectures. }
\label{fig_vecto_particle_times_dynamic_all}
\end{figure}

To confirm that the adaptive mode results in the lowest particle computation time of both scalar and vectorized modes, Fig. \ref{fig_vecto_particle_times_dynamic_all} shows the measured times in the same Maxwellian plasma cases.
In this particular configuration, the plasma remains uniform in the whole domain during the simulation, but the computation times vary depending on the (initial) number of PPC.
As expected, the inversion point between scalar and vectorized modes is located between 10 and 12 PPC.
The adaptive mode 1 is slightly more efficient than mode 2 below 12 PPC because of the slightly more expensive sorting method.
As expected, both modes provide the same performances above 12 PPC.

\begin{figure}
 \centering
 \includegraphics[width=0.49\textwidth]{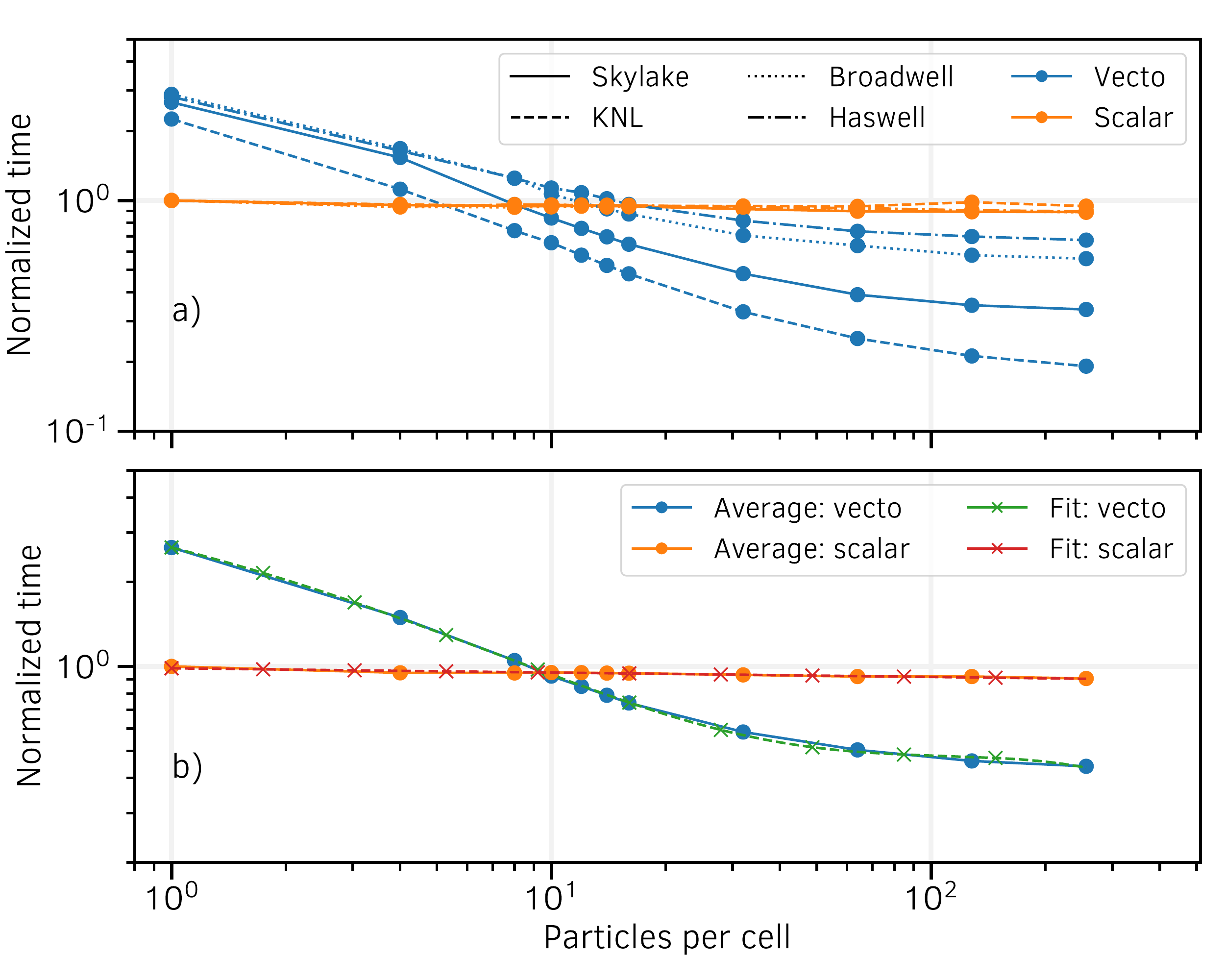}
 \caption{a) Normalized time per particle spent for all
 particle operators in the and vectorized modes with various architectures, and 4th-order interpolation shape functions.
 b) Averages of the curves in panel a), and polynomial regressions. }
\label{fig_vecto_efficiency_o4_all_fit}
\end{figure}

Fig. \ref{fig_vecto_efficiency_o4_all_fit} shows
the normalized times for a 4th order shape factor.
Contrary to the 2nd order case, the difference between all architectures is
more important and the use of a general fitting function is less reliable.
Nevertheless, averages for all architectures and
 polynomial regressions are shown in the same figure as they provide a sufficient estimate of the vectorization speed gain.


\section{Simulation performance benchmarks}
\label{sec:simulation_benchmark}

In this section, the advantages of the adaptive mode will be presented considering three different simulation setups.
The first two are related to laser-plasma interaction at ultra-high intensity, the last one to astrophysics.
All three setups have been chosen as typical of current interests from the plasma simulation community,
and realistic parameters have been chosen for each setup.
In all cases,  the second order interpolation and (Esirkepov) projection was used.
All simulations have run on the Skylake partition of the \textit{Irene Joliot-Curie} supercomputer.

\subsection{Laser Wakefield Acceleration}


Laser wakefield acceleration (LWFA)  consists in accelerating electrons
in the wake of a laser propagating through a low density (transparent) plasma.
A plasma wave is generated in the wake of the laser pulse as a result of the collective response of the electrons
to the electromagnetic field associated to the laser pulse\cite{Malka2002,Esarey2009,Malka2012}.
At large laser intensities, nonlinear effects may lead to a succession of
electron-depleted cavities separated by steep and dense electron shells,
instead of a smooth sinusoidal  wave.
In some specific cases, the cavities look like bubbles, empty of electrons,
and one then speaks of the \textit{bubble regime} of acceleration\cite{Pukhov2002}.
At the back side of the bubble (the front side being the one closest to the laser pulse),
some electrons can be injected in the first half of the bubble and then accelerated forward due to the existence of a strong
negative longitudinal electric field, eventually reaching speed close to that of light.
This method is used in the laboratory to accelerate electrons up to energies at the multi-GeV level over very short distances of a few mm to a few cm.
A strong effort is made to improve the control and quality of the produced electron beams.
It depends on various plasma and laser parameters and this effort strongly relies on massive 3D PIC simulation.


In this Section, we study the impact of the vectorization strategy on LWFA.
To do so, three series (considering 4, 8 and 16 PPC, respectively) of three simulations
(considering the scalar, vectorized and adaptive modes) are presented.
In these simulations, a laser pulse with wavelength $\lambda$
(corresponding to an angular frequency $\omega=2\pi c/\lambda$)
is sent onto a fully ionized hydrogen plasma.
The plasma density profile consists in a long linear ramp (from $x=100\  c/\omega$ to $1280\  c/\omega$)
preceding a plateau at the density $n_0 = 5 \times 10^{-3} n_c$, with $n_c = \epsilon_0 m_e \omega^2/e^2$ the critical density.
The laser pulse, with maximum field strength  $a_0=10$ (in units of $m_e c \omega/e$) is injected from the $x=0$ boundary.
It has a Gaussian temporal profile of $20\pi\ \omega^{-1}$ FWHM (Full Width at Half Maximum)
and a Gaussian transverse spatial profile of waist $24\pi \ c / \omega$.
Its propagation through the plasma is followed up to a distance of $2050\  c/\omega$.
Yet, instead of simulating the full propagation length, which would be too costly, the simulation domain consists in a moving window
sufficiently large to contain the laser and a few wakefield periods and traveling at the laser group velocity.
The overall domain has a dimension of $503 \times 503 \times 503\ (c/\omega)^3$.
It is discretized in $1280 \times 320 \times 320$ cells, corresponding to spatial steps of
$\Delta_x = 0.39\ c / \omega \sim \lambda / 16$ and $\Delta_y = \Delta_z = 0.157\ c / \omega \sim \lambda/4$.
The time step is computed from the CFL condition as $\Delta_t = 0.96 \Delta_{\rm CFL} \simeq 0.31 \omega^{-1}$.
A patch contains $10\times10\times10$ cells, for a total of $128 \times 32 \times 32$ patches.
Only the electron species is considered and an immobile ion background is assumed by
using the charge conserving current deposition scheme (with no Poisson solver at initial time).
This simulation setup was run with the scalar, vectorized and adaptive modes, with 4, 8 and 16 PPC at initialization.
The adaptive mode reconfiguration is done every 50 iterations.
These simulations have run on 96 Skylake processors (48 nodes), corresponding to 2305 cores,
with 1 MPI process per processor and 24 OpenMP threads per MPI process.

\begin{figure}
 \centering
 \includegraphics[width=0.49\textwidth]{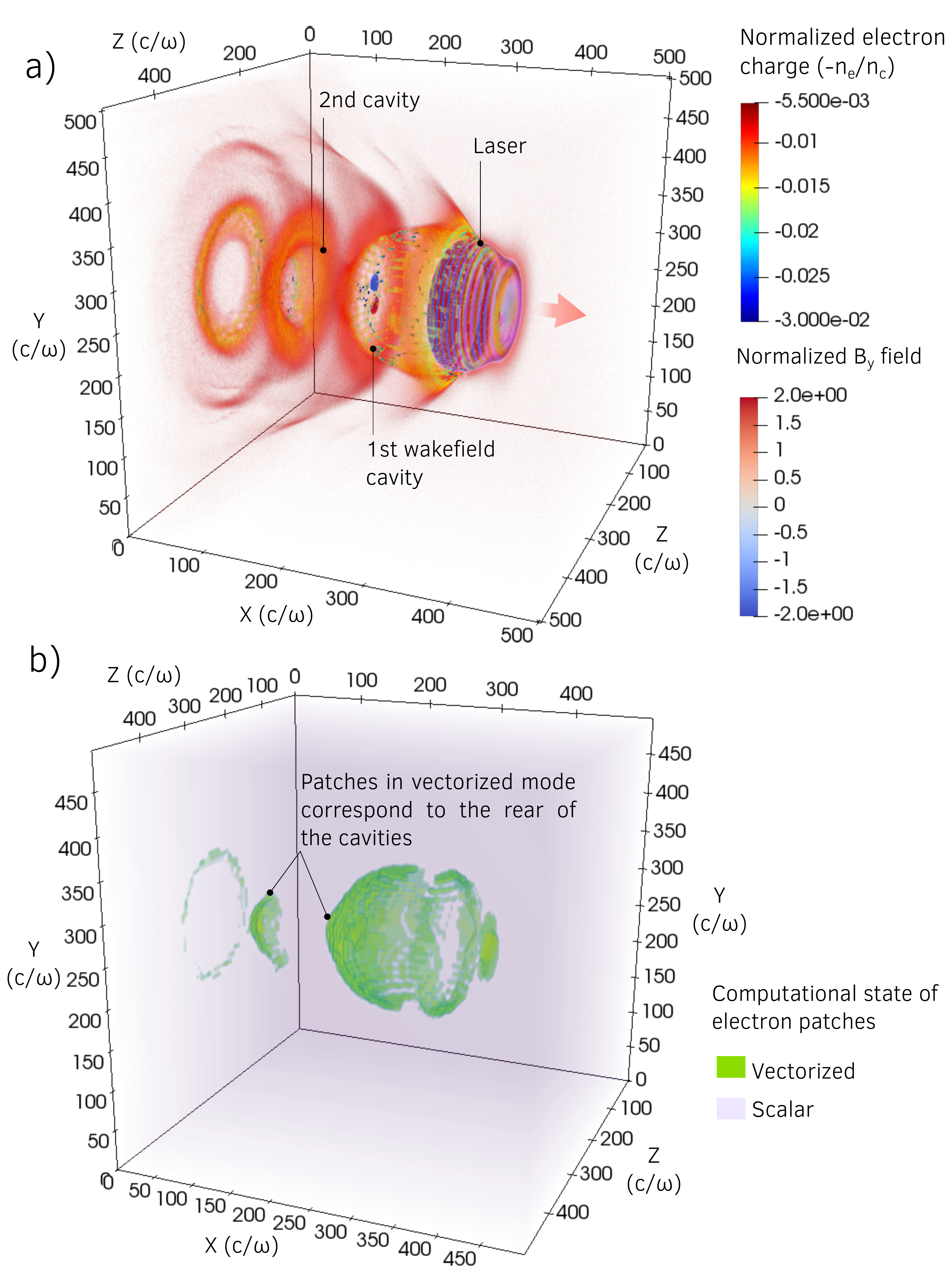}
 \caption{Laser wakefield acceleration.
 a) Volume rendering of the electron charge density (in units of $e n_c$) and laser magnetic field ($B_y$, in units of $m_e \omega/e$),
 at time $t =2770\ \omega^{-1}$.
 b) Patches using a vectorized operator for the electron species at the same time.
An animated version of these quantities can be viewed in the supplementary materials.}
\label{fig_LWFA_3d_ne_vecto_9000}
\end{figure}

\begin{figure}
 \centering
 \includegraphics[width=0.49\textwidth]{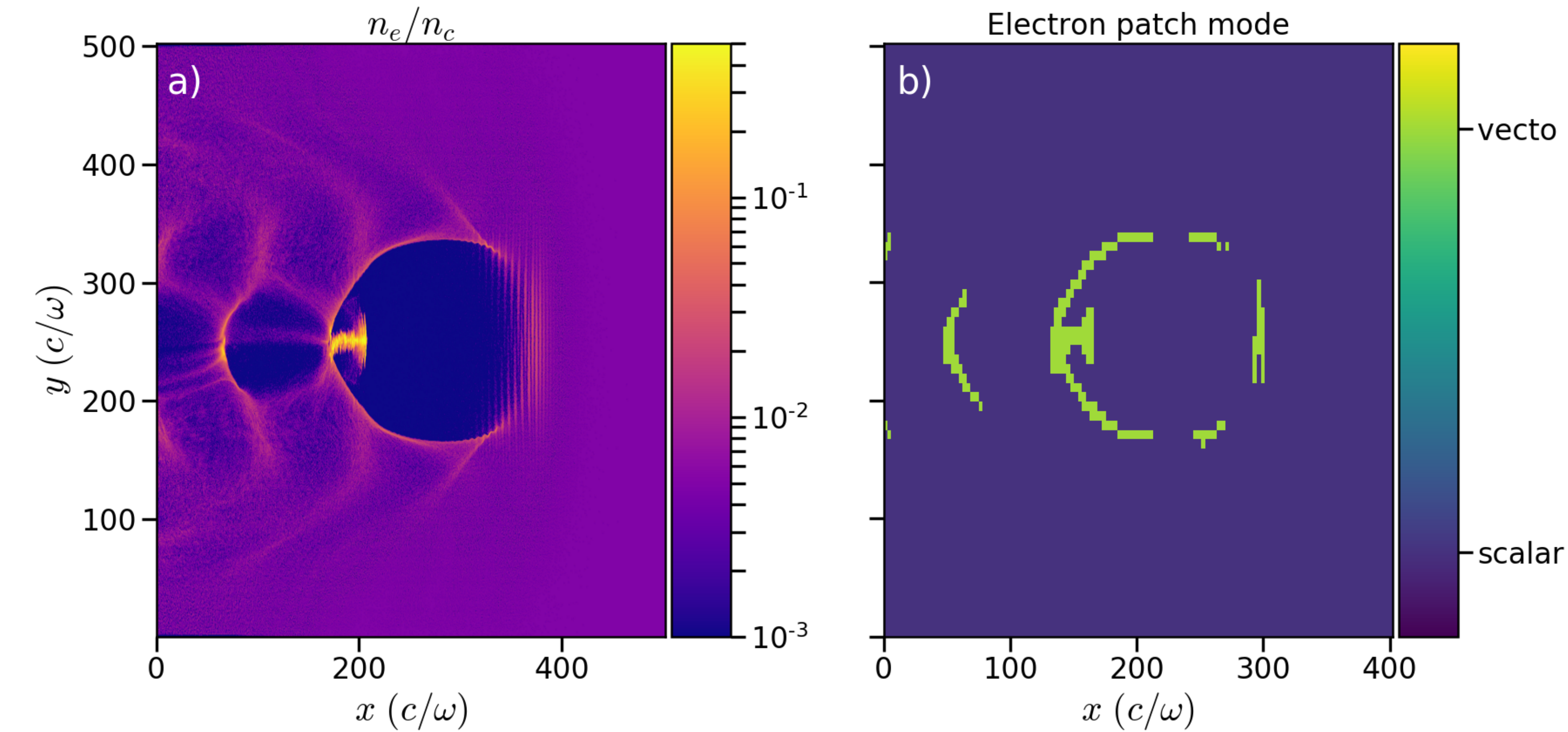}
 \caption{Laser wakefield acceleration. Same as Fig.~\ref{fig_LWFA_3d_ne_vecto_9000}
 but taking a 2D slide at $z = 0.5\ L_z$ ($L_z$ being the domain length in the $z$ direction)
 a) Electron charge $-n_e / n_c$ at time $t =2770\ \omega^{-1}$.
 b) Patches using a vectorized operator for the electron species at the same time.}
\label{fig_LWFA_2d_ne_vecto_9000}
\end{figure}

Figure \ref{fig_LWFA_3d_ne_vecto_9000}a shows a volume rendering of the electron density
illustrating the wakefield cavities surrounded by dense electron layers.
For the reader's convenience, Fig. \ref{fig_LWFA_2d_ne_vecto_9000}a presents a 2D
slice of the electron density taken at $z = 0.5\ L_z$ ($L_z$ being the domain length in the $z$ direction).
At the rear of each cavity, very high density electron bunches are accelerated by the strong charge separation electric field.
Those beams have a density that can be several orders of magnitude higher than the initial plasma density,
which translates in a large load imbalance.
In particular, patches in these high-density regions see their average number of particles per cell largely exceeding the initial one,
and will thus benefit most of the vectorized operator.
Figure \ref{fig_LWFA_3d_ne_vecto_9000}b (see also Fig.~\ref{fig_LWFA_2d_ne_vecto_9000}b) highlights the regions
where the adaptive mode has switched to vectorized operators.
As expected, these regions corresponds to the patches containing a large number of PPC,
such as the rear side of the wakefield cavities (containing the electron beams)
and a thin circle around the first cavity at $x = 200\ c/\omega$.
Let us note an additional advantage of the adaptive mode which helps mitigating the load imbalance at the node level
as patches holding many particles can be treated more efficiently than those holding only few of them.



\begin{figure}
 \centering
 \includegraphics[width=0.49\textwidth]{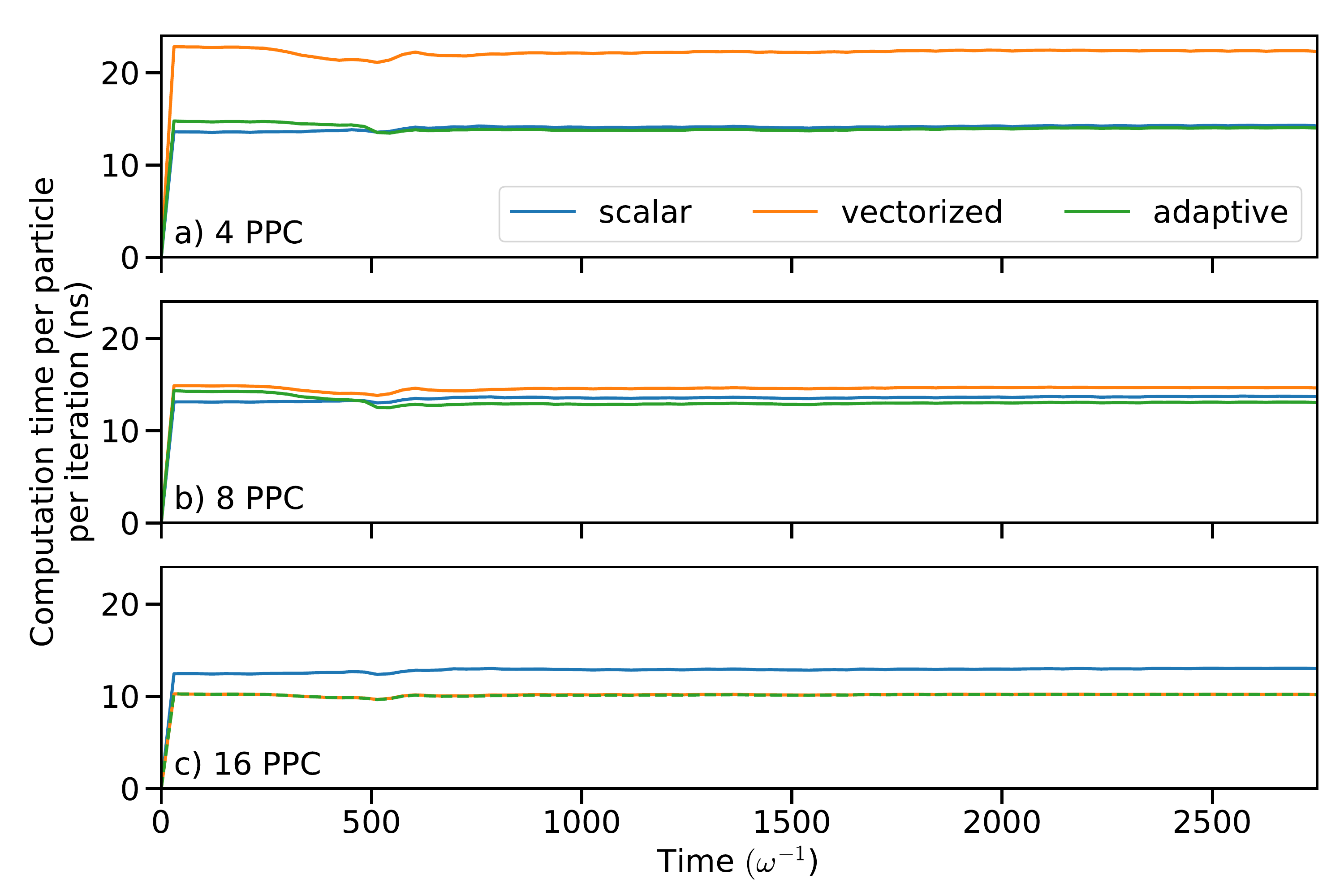}
 \caption{Laser wakefield acceleration.
 Temporal evolution of the mean particle computation
  time (only in the particle operators) spent per
 particle per iteration at 4, 8 and 16 PPC (respectively figures a, b and c
 for the scalar, vectorized and adaptive modes.}
\label{fig_LWFA_particle_time}
\end{figure}

Figure \ref{fig_LWFA_particle_time} presents the temporal evolution of the computation (node) time per particle and iteration
considering 4, 8 and 16 PPC (panels a, b and c, respectively).
One recovers that when using few particles per cell (4 PPC in panel a), the scalar operator is the most efficient one,
while considering a larger number of particles per cells (16 PPC in panel c), the vectorized one is more interesting.
Importantly, the adaptive mode allows to select the optimal operator for all three panels and provides the most efficient approach.

Overall, considering 4 PPC, the computation time spent in the particle operators is close to $680$ s for both the scalar and adaptive modes,
and of $1080$ s for the vectorized one.
As most of the simulation box contains few PPC, the adaptive mode selects adequately the scalar operator.

With 8 PPC, the computation times for all three modes are similar,
equal to 1300 s (scalar), 1400 s (vectorized) and $1260$ s (adaptive).
This number of PPC is indeed close to what was referred to as the inversion point in Sec.~\ref{sec:vecto_efficiency}.

With 16 PPC, the vectorized mode is the most efficient with a computation time of $1950$ s,
while the scalar mode is significantly slower with a particle computation time of $2500$ s.
The adaptive mode hence selects adequately the vectorized operator leading to
the same time of $\sim 1950$ s, see also Fig. \ref{fig_LWFA_particle_time}c.

Let us finally note that the computation time per particle per iteration decreases
with the number of PPC using the adaptive mode while it is barely
sensitive to the number of PPC considering the scalar one.
At the modest number of 16 PPC, the vectorized operator already allows to
decrease the computation time by more than 20\% with respect to the scalar one.
Finally, for all cases, the time allocated to the adaptive reconfiguration process is well below 1\% of the simulation time.


\subsection{Laser interaction with a solid-density thin foil}


Laser interaction with high-density ($n_0 \gg n_c$) plasmas created by irradiating solid-density foils
is at the center of various experimental and theoretical investigations by the laser-plasma community.
These studies are motivated by the broad range of physical mechanisms and potential applications
of this kind of interaction, ranging from electron and ion acceleration, new radiation sources
(from THz to XUV and $\gamma$), to the possibility to address strong field quantum electrodynamicss effects
\cite{daido2012,macchi2013,dipiazza2012}.

In this Section, we illustrate the impact of the vectorization strategy on the simulation of such a high-density target irradiated by an ultra-intense laser pulse.
To do so, three simulations using either the scalar, vectorized or adaptive operators are reported.
In these simulations, a laser pulse  with wavelength $\lambda$ (corresponding to an angular frequency $\omega = 2\pi c/\lambda$)
is focused at normal incidence onto a carbon foil located at $\sim 37.7\ c/\omega$ ($6\lambda$) from the $x=0$ boundary.
The carbon foil is a fully-ionized plasma which density increases from 0 to its maximum $n_0 = 492\ n_c$ linearly over $\sim 12.6\  c/\omega$ ($2\lambda$)
(this ramp mimics a pre-plasma) then forms a plateau with thickness $\sim 12.6\ c/\omega$ ($2\lambda$).
The foil density is otherwise uniform over the full simulation domain in the transverse ($y$ and $z$) directions.
At initialization, both carbon ions and electrons have the same uniform temperature of 1 keV.
The laser pulse, with maximum field strength $a_0=100$ (in units of $m_e c \omega/e$), is injected from
the $x=0$ boundary. It has a fourth-order hyper-Gaussian temporal profile of FWHM $\sim 188.5\ \omega^{-1}$ ($30\lambda/c$)
and a transverse Gaussian profile with waist $\sim 12.6\ c/\omega$ ($2\lambda$).
It is focused at the front of the preplasma ($x=6\ \lambda$) and at the center of the simulation box in the $y$ and $z$ directions.
The simulation lasts for 100 laser periods ($\lambda/c$), the time to fully complete the laser interaction.
The simulation domain extends over $\sim 100 \times 67 \times 67 (c/\omega)^3$ (approximately $16\lambda \times 11\lambda \times 11\lambda$)
discretized in $1024 \times 256 \times 256$ cells, corresponding to a spatial resolution $\Delta x = \lambda / 64 \simeq 0.10 c/\omega$
and $\Delta y = \Delta z = \lambda / 24 \simeq 0.26 c/\omega$, and the time step is $\Delta_t = 0.96 \Delta_{\rm CFL} \sim  0.083\  \omega^{-1}$.
Cells containing plasma are initialized with 32 randomly-distributed PPC and
the simulation domain is decomposed into $128\times 32 \times 32$ patches,
each patch containing $8\times8\times8$ cells.
Three simulations have been run considering the scalar, vectorized and adaptive modes, respectively.
For the latter, the adaptive mode reconfiguration is done every 8 iterations.
These simulations run on 64 Skylake processors (32 nodes, 3072 cores) with 1 MPI process per processor
and 24 OpenMP threads per MPI process.

\begin{figure}
 \centering
 \includegraphics[width=0.49\textwidth]{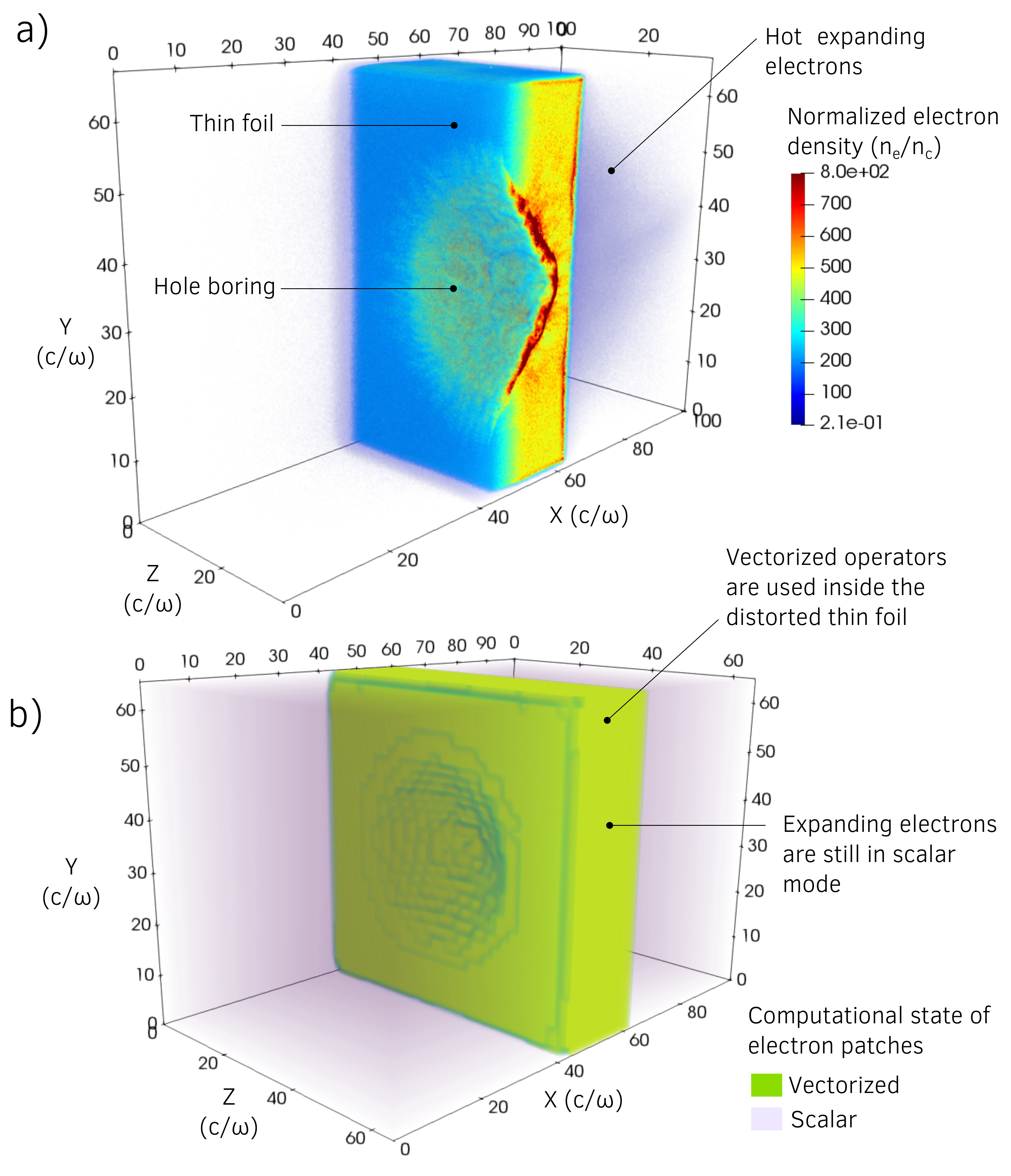}
 \caption{Laser over-dense foil interaction.
 a) Volume rendering of the normalized electron density  $n_e / n_c$ at time $t = 537\ \omega^{-1}$.
 Only half of the target (subset $0 < z \leq 0.5 L_z$)  is shown.
 The cross-section highlights the inner target structure in addition to the outer target distortion effects.
 b) Patches using a vectorized operator (adaptive mode) for the electron species at the same time.
 An animated version of these quantities can be viewed in the supplementary materials.}
\label{fig_TF_ne_vecto_3D}
\end{figure}

\begin{figure}
\centering
\includegraphics[width=0.49\textwidth]{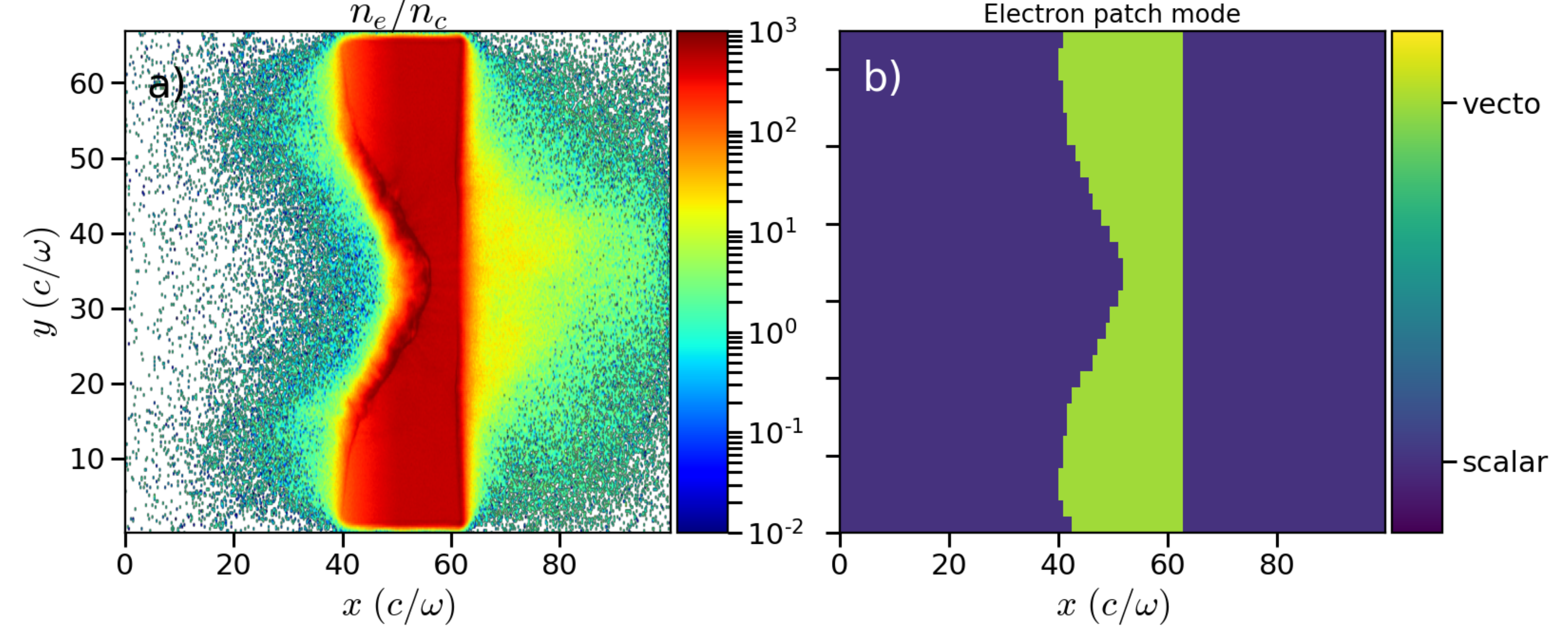}
\caption{Laser over-dense foil interaction.
a) Slice at $z = 0.5 L_z$ of the electron density $n_e / n_c$ at time $t = 475 \omega^{-1}$
corresponding to the end of the laser interaction.
b) Patches using a vectorized operator (adaptive mode) for the electron species at the same time..}
\label{fig_TF_ne_vecto}
\end{figure}

Figure \ref{fig_TF_ne_vecto_3D}a illustrates the deformation of the foil as it is irradiated by the ultra-intense laser pulse.
Indeed, the overdense (i.e. with density $n_0>n_c$) plasma is opaque to the laser light which is thus reflected at the foil's surface.
As the laser pulse bounces off the target, it exerts a strong (radiation) pressure onto its surface which is pushed inward,
a process known as {\it hole boring} and highlighted in Fig.~\ref{fig_TF_ne_vecto_3D}a.
At the same time, the laser plasma interaction in the pre-plasma at the target front side leads to the copious production
of relativistic electrons that propagate throughout the foil, and eventually escape at its back as a hot, low density, electron gas.
Also illustrated in Fig.~\ref{fig_TF_ne_vecto_3D}a, this tenuous electron plasma escaping from the target is better illustrated in
Fig. \ref{fig_TF_ne_vecto}a, showing a 2D slice (at $z = 0.5 L_z$) of the electron density in logarithmic scale.

Figure \ref{fig_TF_ne_vecto_3D}b presents, for the simulation in adaptive mode, the distribution of patches relying on
vectorized operators. Interestingly, these patches are located where the particle density is high, that in the region
corresponding to the initial target location minus the front side hole-boring region that has been depleted of particles.
Note also that patches located in the region at the back of the target, where hot electrons are escaping, use the scalar
operator as the hot electron gas is tenuous and thus described by only few PPC.


\begin{figure}
 \centering
 \includegraphics[width=0.49\textwidth]{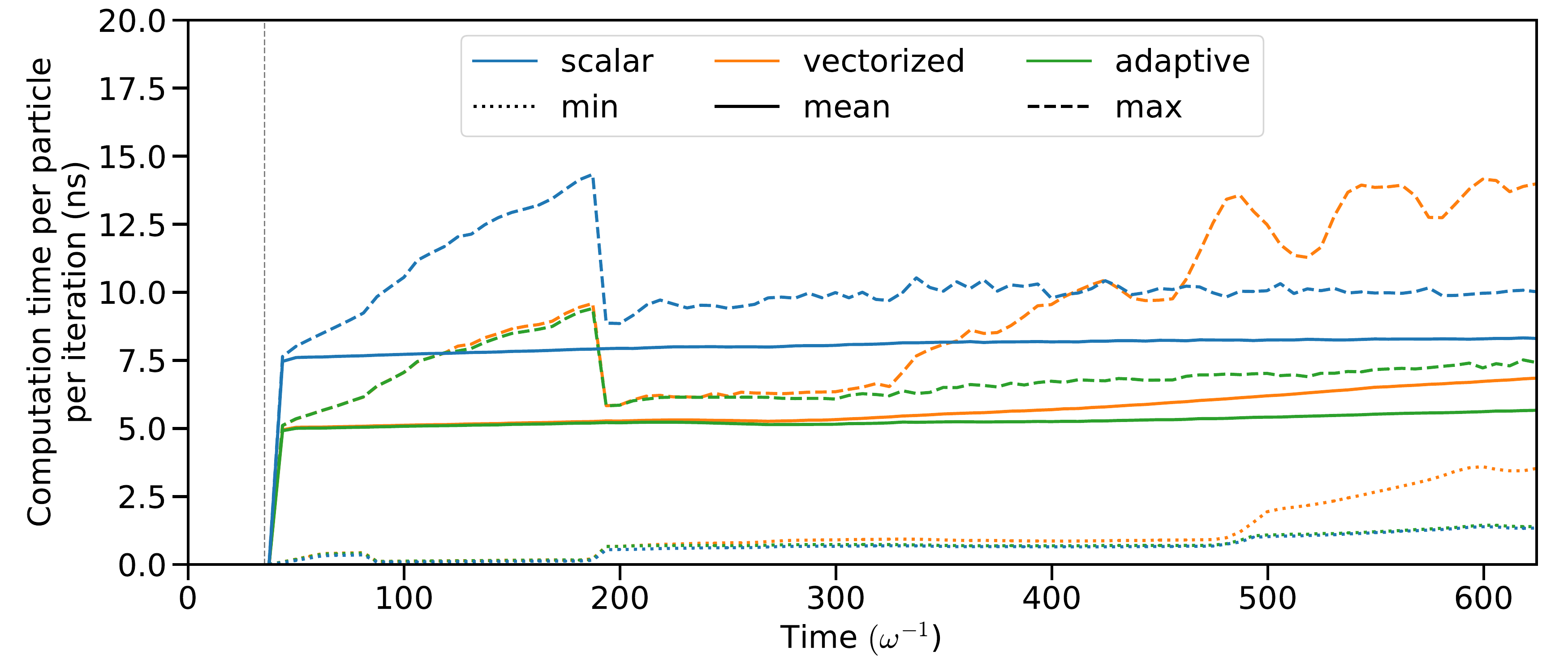}
 \caption{Laser over-dense foil interaction. Temporal evolution of the computation
 particle time (only in the particle operators) spent per
 particle per iteration for the scalar, vectorized and adaptive modes.
 Among all MPI processes, average, minimum and maximum times are shown.
 Note that the time acquisition is started at $t=37 \omega^{-1}$ when
the laser strikes the target.}
\label{fig_TF_particle_time}
\end{figure}

Figure \ref{fig_TF_particle_time} presents the temporal evolution of the computation (node) time per particle and iteration.
The mean value (solid line) shows that for these simulations, the adaptive mode is the one that provides the most efficient treatment
over the full simulation.
In addition to the mean value, the computation (node) time per particle and iteration was also computed for each MPI task separately,
and the minimum and maximum values reported in Fig. \ref{fig_TF_particle_time}.
The maximum value is particularly interesting as it refers to the computation time on the least efficient MPI task.
Following this value in time allows to see how the adaptive mode adapt to each phase of the physics process.
At early times, the dense target is associated to a large number of PPC,
the vectorized operator is the most efficient one, and adaptive mode adequately select it.
At later times, the electron population expends, its density decreases and more and more patches with few PPC are generated.
As a result, the vectorized mode becomes less and less efficient and, at $t\sim 400\omega^{-1}$, the scalar operator becomes more interesting.
The adaptive mode thus eventually selects the scalar operator, and throughout the simulation, the adaptive mode is the
one that proves the most efficient.

Overall, after 7550 iterations, the computation time spent in the particle operators is
912 s with the scalar mode,
647 s with the vectorized mode
and 604 s with the adaptive mode.
The adaptive mode thus allows to reduce the simulation time by $\sim 34\%$,
and for this case, the overhead due to the adaptive reconfiguration of is also below the percent.


\subsection{Mildly-relativistic collisionless shock}


Ubiquitous in astrophysics, collisionless shocks have been identified as one of the
major sources of high-energy particle and radiation in the Universe~\cite{KirkDuffy1999},
and, as such, have been the focus of numerous PIC simulations over the last decade~\cite{spitkovsky2008,sironi2013,plotnikov2018}.
Collisionless shocks can form during the interpenetration of two colliding plasmas.
In the absence of external magnetic field, the Weibel instability~\cite{weibel1959} provides the dissipation mechanism necessary to shock formation.
This instability quickly grows in the overlapping plasma region (see, e.g. \cite{grassi2017}), and leads to the formation of current filaments associated with a strong magnetic field perturbation.
At the end of the linear phase, the magnetic and current filaments distort into a region of electromagnetic turbulence,
decelerating and transversely heating the flow's particles, ultimately leading to their isotropization and thermalization.
This leads to a pile-up of the particle in the turbulent region during which both the plasma density and pressure increase up to the formation of a shock front.

To illustrate this process and the impact of adaptive vectorization on its simulation,
three series (considering either 4, 8 or 32 PPC)
of three simulations (using the scalar, vectorized and adaptive modes) are presented.
In these simulations, two counter-propagating electron-positron plasma flows are
initialized each filling half of the simulation domain (in the $x$-direction).
Both flows, with density $n_0$, have opposite drift velocity $\pm 0.9~c$ (in the $x$-direction),
corresponding to a Lorentz factor $\gamma_0 = 2.3$, so that they collide at the center of the 3D simulation domain.
The domain size is $300 \times 28.5 \times 28.5 \ (c/ \omega)^3$,
with $\omega=\sqrt{e^2 n_0/(m_e\epsilon_0)}$ the electron plasma frequency associated to the initial flow density $n_0$.
The cell sizes were set to $\Delta_x \simeq 0.11 \ c/\omega$ and $\Delta_y = \Delta_z \simeq 0.15 \ c / \omega$,
and the time step to $\Delta_t = 0.95 \Delta_{\rm CFL}$.
This ensures a good resolution of the relativistic electron skin-depth $d_{e,{\rm rel}} = \sqrt{\gamma_0} c/\omega \simeq 1.5\,c/\omega$
and thus of the Weibel filaments.
The simulation lasts $100\  \omega^{-1}$.
Each patch contains $8 \times 8 \times 8$ cells, initialized with either 4, 8 or 32 randomly-distributed PPC.
The adaptive mode reconfiguration is done every 8 iterations.
These simulations have been run on 64 Skylake processors (32 nodes), corresponding to 1536 cores.

\begin{figure} 
 \centering
  \includegraphics[width=0.49\textwidth]{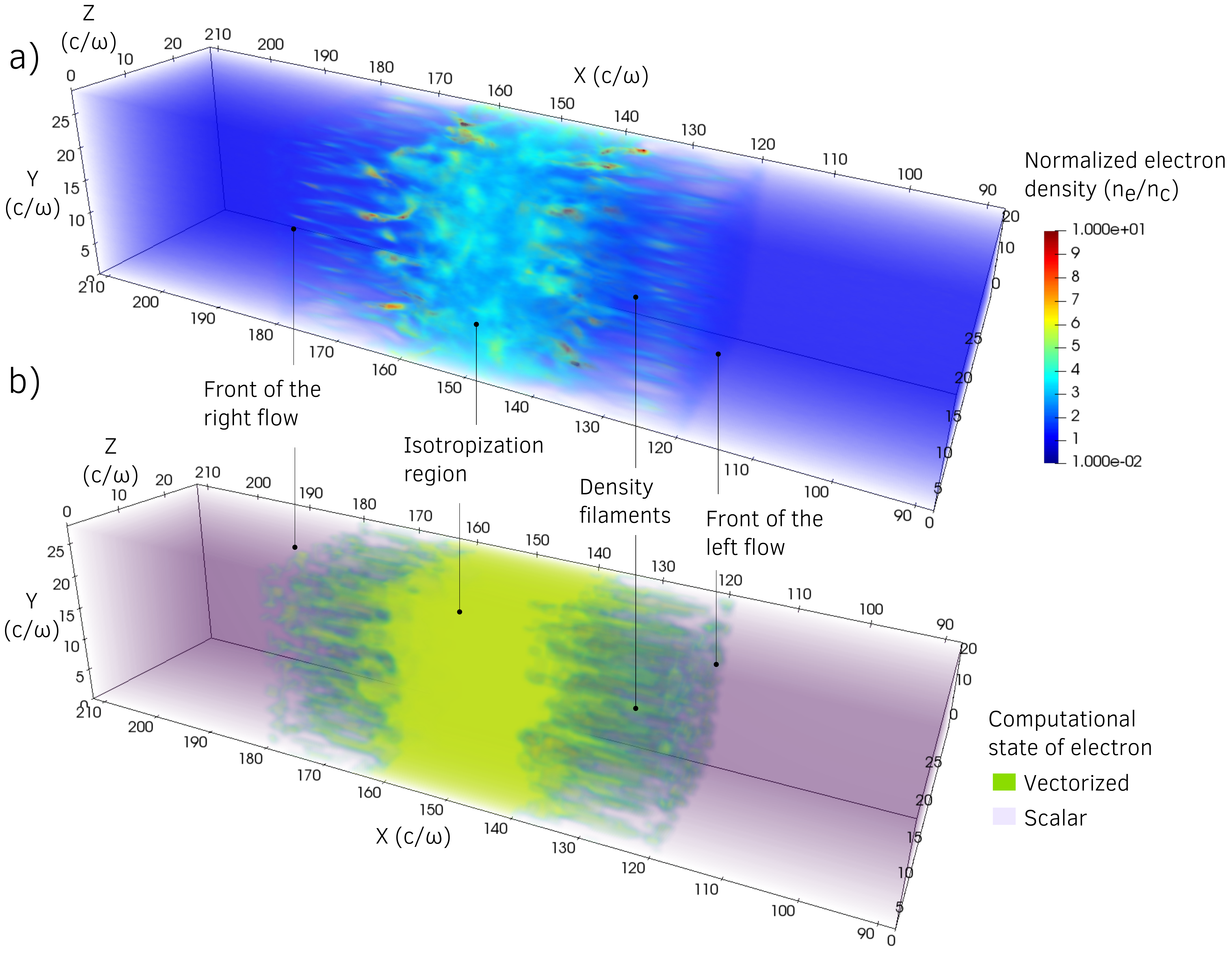}
 \caption{a) Volume rendering of the normalized electron density
 $n_e / n_c$ at time $t =34\ \omega^{-1}$ after the beginning of the collision.
 b) Patches in vectorized mode for the electron species at the same time.
  An animated version of these quantities can be viewed in the supplementary materials.}
\label{fig_Weibel_ne_vecto_3D}
\end{figure}

\begin{figure} 
\centering
\includegraphics[width=0.49\textwidth]{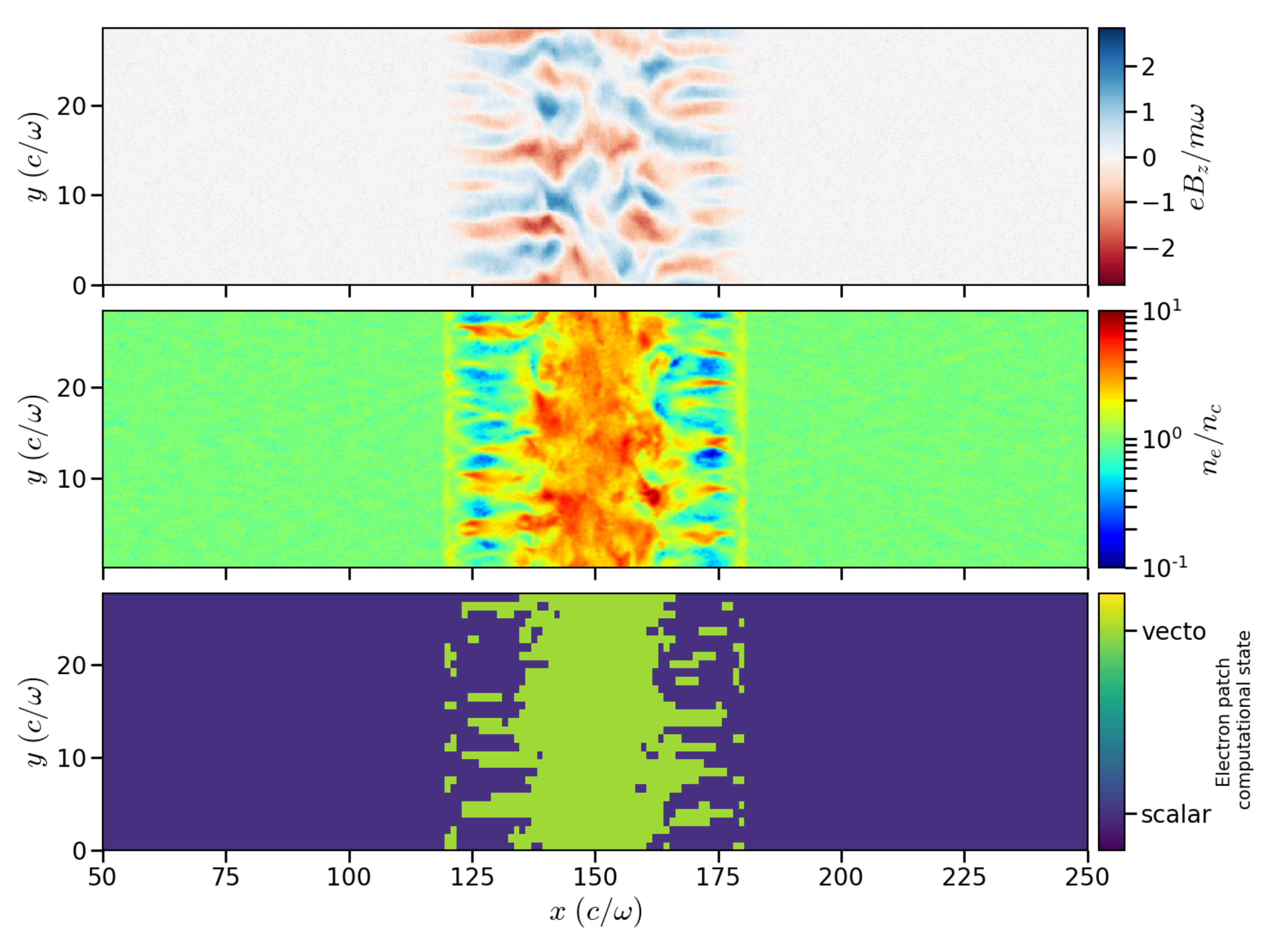}
\caption{a) Slice of the transverse normalized magnetic field $B_z / B_0$
during the plasma flow collision at $z = 0.5 L_z$, $L_z$ being the domain length
in the $z$ direction at time $t = 34\ \omega^{-1}$.
b) Slice of the normalized electron density $n_e / n_c$
 at time $t = 34\ \omega^{-1}$ and $z = 0.5 L_z$.
c) Computational mode (scalar or vectorized) of the electron species for each
 patch in the slice $z = 0.5 L_z$ at $t = 34\ \omega^{-1}$.}
\label{fig_Weibel_bz_ne_vecto}
\end{figure}


Figure \ref{fig_Weibel_ne_vecto_3D}a shows a 3D volume rendering of the electron density at an early stage of the interaction ($t = 34\omega^{-1}$).
The Weibel filamentation region is clearly illustrated as well as the on-set of turbulence in the central region.
Figure \ref{fig_Weibel_ne_vecto_3D}b shows, at the same time, the distribution of patches for which the adaptive mode switched to the vectorized operator.
It is clear that these patches are located in the high-density regions at the position of the Weibel filaments,
as well as in the central region where the density increases by a factor nearly of $\times 4$, as expected for a fully formed 3D shock.
For the reader's convenience, a 2D-slice taken at $z = 0.5 Lz$ is also presented in Fig.~\ref{fig_Weibel_bz_ne_vecto}.
In panel a is also presented the magnetic field structure characteristic of the Weibel instability and its latter more turbulent state in the central region.



\begin{figure}
 \centering
 \includegraphics[width=0.49\textwidth]{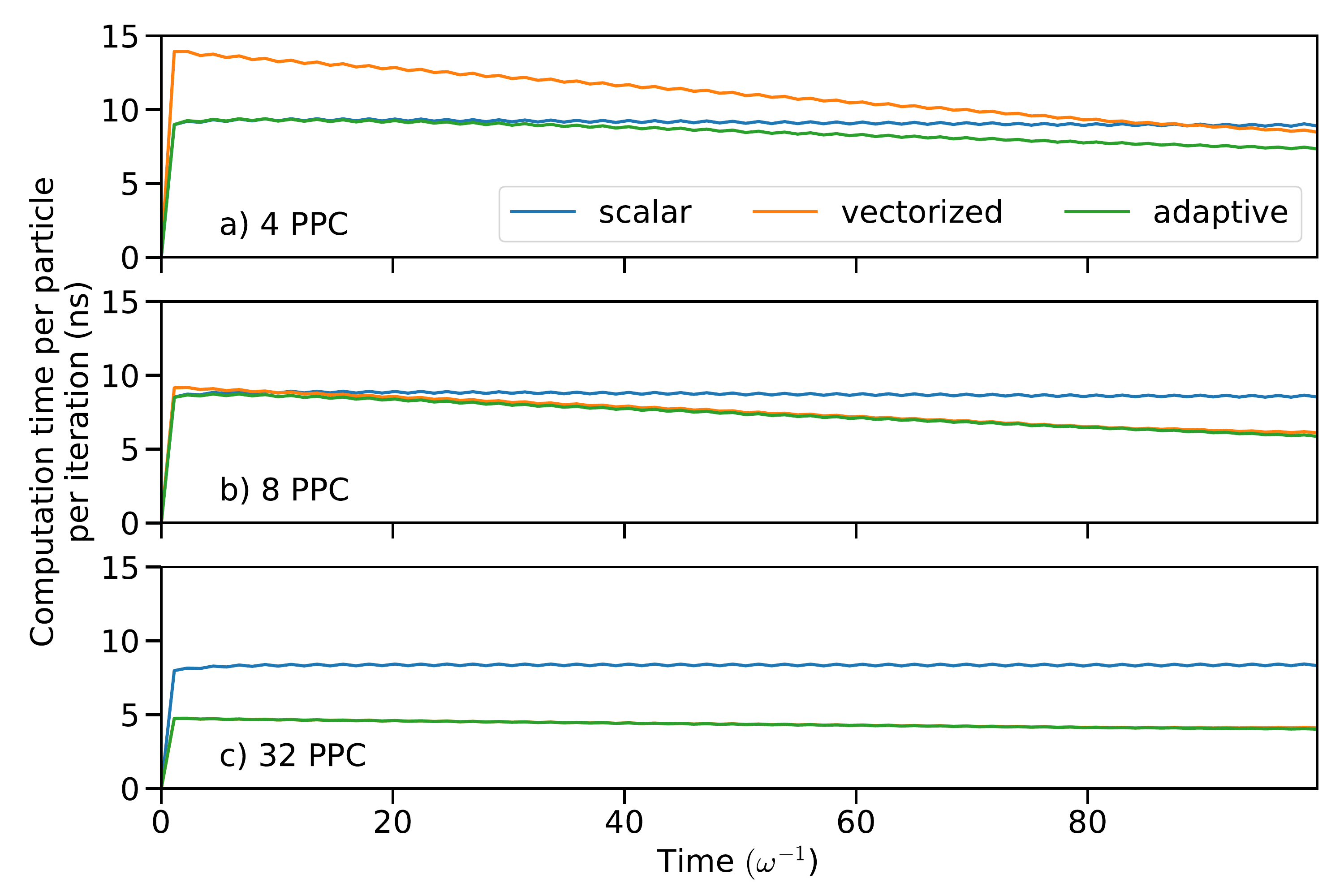}
 \caption{For the collisionless shock simulations: temporal evolution of the mean particle computation
  time (only in the particle operators) spent per
 particle per iteration at 4, 8 and 32 PPC (respectively figures a, b and c)
 for the scalar, vectorized and adaptive modes.}
\label{fig_Weibel_particle_time}
\end{figure}

Figure \ref{fig_Weibel_particle_time} provides the detailed evolution of the computation
(node) time per particle per iteration at respectively 4, 8 and 32 PPC (panels a, b, c, respectively).
As expected, at 4 PPC, the scalar operator is the most efficient and the adaptive mode adequately selects it,
leading to similar computation time.
As the simulation goes on, particles starts piling up in the overlapping region at the center of the  simulation domain.
The effective number of PPC in the central patches increases and the vectorized operator becomes more and more interesting with respect to the scalar one.
The adaptive mode benefits from this speed-up by selecting the vectorized operator wherever it allows from improved efficiency.
For 8 and 32 PPC, the vectorized operator and the adaptive mode lead to the highest efficiency.

Overall, at 4 PPC, the computation time spent in particle operators is for the full simulation (1515 iterations) of 433 s, 534 s and 405 s in the scalar, vectorized and adaptive mode, respectively.
Even with such a small number of PPC, the adaptive approach allows for 6\% gain in efficiency with respect to the scalar mode.
At 8 PPC, the particle computation time is of 826 s, 723 s and 709 s for the  the scalar, vectorized and adaptive mode, respectively.
The gain in efficiency thus increases to 14\%.
Finally, at 32 PPC, the particle computational time is of 3160 s for the scalar mode and reduced to 1660 s for both the vectorized and adaptive modes.
In this case, the gain in efficiency due to the vectorized operators is of 47\%, that corresponds to a speed-up of almost $\times 2$ for this configuration.

Last, we note that, in this configuration again, the time per particle per iteration decreases with the number of particles.
In addition, the overhead due to the adaptive reconfiguration remains for all cases below 1\% of the full simulation time.


\section{Conclusion}
\label{sec:conclusion}


The new vectorized particle operators implemented in \Smilei rely on an
optimized cycle sort.
It sorts particles by dual cell (particles with the same primal index
are contiguous in memory) at all times.
It takes advantage from the low number of particles changing cells between time steps and from the fact that there are several particles in each cell.
The number of particle copies is minimized and the algorithm complexity is reduced to $\mathcal{O}\left(N_{part}\right)$
by keeping track of the cell locations, as in a counting sort.


The interpolation operator has been efficiently vectorized thanks to this sorting method
that avoids random memory access and facilitates data reuse.
Although the pusher was already efficiently vectorized using the particles' structure of arrays, it is now applied more efficiently on particle groups instead of the full arrays.
The Esirkepov projection, being hardly vectorizable in its naive implementation due to concurrent memory access, has taken advantage
of the new cycle sort and thus shown an efficient vectorization.
The presented method computes particles by groups of 8 and uses temporary buffers sized accordingly (a reduction step is necessary to update eventually the main arrays).
Clusters of 4 cells are considered to limit the memory footprint
and the number of reductions.
An improved efficiency of the vectorized operators is obtained,
compared to their original scalar implementation, when the number of particles per cell is
sufficiently large, generally above 8 particles per cell.
This threshold depends on the processor
architecture (vector instruction set) and the order of the interpolation shape functions.
In all cases, the vectorized operators, combined with the cycle sort, significantly
speed up the particle processing when the number of particles per cell is several
multiples of the vector register length.


But when the number of particles per cell is lower than the vector register length, the vectorized operators become less efficient than their scalar counterparts.
This issue is addressed by using a \textit{adaptive mode} able to pick locally (each patch, each species) and dynamically (every number of time steps) the most efficient version.
Simulations presenting a strong imbalance
in the number of particles per cell contain both vectorized and 
scalar patches, depending on their load. If the plasma evolves,
the mode of each patch changes accordingly.
This adaptive approach results in a simulation cost
equal to or lower than the best mode (scalar or vectorized).
The adaptive reconfiguration overhead appears negligible.
The optimal scenario corresponds, as expected, to a fully
vectorized simulation, but this is not suitable for all physical
cases, hence the adaptive approach.

This adaptive mode does not require any input from the user
as the algorithm detects automatically which operators to pick.
However, the implementation is based on empirical, 
architecture-dependent metrics, and should be reevaluated on other
processor types for optimal performances.
Fortunately, several architectures can
share a similar behavior and it is possible to build common
approximate metrics.
The order of the interpolation shape functions,
the MPI/OpenMP ratio, the compiler version, or other parameters
may also modify these results.
In the future, an automated analysis could be performed by the code at
initialization to compute the most suitable metrics.
For large scale simulations, this evaluation would represent a negligible cost.


There is, for the moment, no overlapping strategy between computation and
communications.
This constitutes a next development axis that would enhance the benefits brought
by this adaptive strategy.
A better integration of the dynamic load balancing with the adaptive vectorization mode constitutes a second possible improvement: they are not
coupled even if they can both contribute separately to the simulation efficiency.
For instance, they do not share the metrics used to estimate the particle computation time.


\section{Acknowledgements} \label{sec:Acknowledgements}

The authors are grateful to M. Haefele, J. Bigot, P. Kestener,
A. Durocher, O. Iffrig, V. Soni and H. Vincenti for fruitful discussions.
This  work  was  granted  access  to  the  HPC  resources of TGCC/CINES under
the allocation 2017 - A0020607484 and \textit{Grand Challenge} ``Irene'' 2018
project gch0313 made by GENCI.
The authors are grateful to the TGCC and CINES engineers for their support.
Access to the KNL cluster Frioul was granted via the Cellule de Veille
technologique.
The authors acknowledge the European EoCoE project for sharing HPC resources
on the supercomputer Jureca.
The authors thank engineers of the LLR HPC clusters for resources and help.
The authors are thankful to M. Mancip for his help in rendering 3D images on
the Mandelbrot cluster and the Mur d'Image.


\appendix
\section{Compilation}
\label{compilation}

There are 4 different clusters used in this article.
Each of them is equipped with processors of
different Intel architectures that represent the most used with \Smilei:
\begin{itemize}
    \item Jureca supercomputer: 2 x Haswell node (Intel Xeon E5-2680 v3, 12 cores)
    \item Tornado supercomputer: 2 x Broadwell node (Intel Xeon CPU E5-2697 v4, 16 cores, 2.3 GHz)
    \item Frioul supercomputer: Knights Landing (KNL) node (Intel Xeon Phi 7250, 68 cores, 1.4 Ghz)
    \item Irene Joliot-Curie supercomputer: 2 x Skylake node (Intel Skylake 8168, 24 cores, 2.7 - 1.9 Ghz)
\end{itemize}
On each of them, the code is compiled with the following versions:
\begin{itemize}
    \item Intel compiler 18.0.1.163, IntelMPI 18.0.1.163
    \item Intel compiler 17.3.191, OpenMPI 1.6.5
    \item Intel compiler 18.0.1.163, IntelMPI 18.0.1.163
    \item Intel compiler 18.0.1.163, IntelMPI 18.0.1.163
\end{itemize}
The most recent architecture is Skylake and uses the extended \textsc{AVX512} vector
instruction set coming from the Xeon Phi family (including KNL).
It has the larger vector size able to treat 8 double precision floats in a
single instruction.
The Skylake architecture can handle by legacy the \textsc{AVX2} instruction set
inherited from the Haswell and Broadwell processors.
The Intel Turboboost technology allows the processor to adjust the core frequency
to the total number of used cores and the required instruction set.
Regarding the Skylake processor used in this article, the base frequency without
vectorization is 2.7 GHz, 2.3 GHz for \textsc{AVX2} and 1.9 GHz for \textsc{AVX512}.

The code is compiled with the most advanced architecture vectorization flags, i.e.
\textsc{-xCORE-AVX2} on Haswell and Broadwell, \textsc{-xMIC-AVX512} on KNL and
\textsc{-xCOMMON-AVX512} on Skylake.
The flag \textsc{-xCORE-AVX2} can also be used on KNL and Skylake to test the code
with the \textsc{AVX2} instruction set that limit the vector register size to
256 bit (4 double precisions float).
These flags are completed by \textsc{-O3 -ip -ipo -inline-factor=1000 -fno-alias}
for best performance.
The KNL cluster is configured in Quadrant cache mode.
On KNL, OpenMP is used to keep the 64 cores busy among the 68 available.
The remaining cores are let alone for the system.
Hyperthreading is not activated.
For the other types of processors, we use all available cores.

\section{Cycle sort}
\label{ap_cycle_sort}

\begin{algorithm}
\DontPrintSemicolon
\KwData{
\\$Particles$: array of unsorted particles.
\\$CellKeys$: array of the cell indexes of the particles.
\\$Npart$: number of particles.
}
\KwResult{$Particles$: array of sorted particles}
\Begin{
\textcolor{blue}{\tcp{Loop on particles}}
 \For{$cycleStart \in range(Npart-2)$}{
     $cell\_dest \longleftarrow CellKeys[cycleStart]$\;
     $ip\_dest \longleftarrow cycleStart$\;
     \textcolor{blue}{\tcp{Compute the destination}}
     \For{$i\leftarrow cycleStart+1$ \KwTo $Npart-1$}{
        \If{$CellKeys[i] < cell\_dest$}{
            $ip\_dest\mathrel{+}= 1$\;
        }
     }
     \If{$ip\_dest== cycleStart$}{
         \textcolor{blue}{\tcp{Particle already well placed}}
         $\mathbf{continue}$\;
     }
     \textcolor{blue}{\tcp{Do not swap twins}}
     \While{$CellKeys[ip\_dest]\mathrel{=}=\ cell\_dest$}{
         $ip\_dest\mathrel{+}= 1$\;
     }
     $Cycle.resize(0)$\;
     $Cycle.push\_back(cycleStart)$\;
     $Cycle.push\_back(ip\_dest)$\;
     \textcolor{blue}{\tcp{Build a cycle}}
     \While{$ip\_dest\ \mathrel{!}=\ cycleStart$}{
         $cycleStart\longleftarrow ip\_dest$\;
         $cell\_dest \longleftarrow CellKeys[cycleStart]$\;
         \For{$i\leftarrow cycleStart+1$ \KwTo $Npart-1$}{
             \If{$CellKeys[i] < cell\_dest$}{
                 $ip\_dest\mathrel{+}= 1$\;
             }
         }
         \While{$CellKeys[ip\_dest]\mathrel{=}=\ cell\_dest$}{
         $ip\_dest\mathrel{+}= 1$\;
         }
          $Cycle.push\_back(ip\_dest)$\;
     }
     \textcolor{blue}{\tcp{Proceed to the swap}}
     $Ptemp\longleftarrow Particles[Cycle[0]]$\;
     \For{$i \leftarrow Cycle.size()-1$ \KwTo 2}{
         $Particles[Cycle[i]]\longleftarrow Particles[Cycle[i-1]]$\;
     }
     $Particles[Cycle[1]]\longleftarrow Ptemp$\;
 }

\KwRet  $Particles$

}
\caption{Cycle Sort.}
 \label{cyclesort}
\end{algorithm}

\bibliography{bibliography}{}
\bibliographystyle{unsrt}

\end{document}